\documentclass{aa}

\usepackage{ifpdf}
\usepackage{makeidx}
\usepackage{amssymb}
\usepackage{amsmath}
\usepackage{eurosym}
\usepackage{amsfonts}

\usepackage{graphicx,natbib}

\newcommand{\loe}{\stackrel{<}{\sim}}

\newcommand{\etal}{{\it et al.}}
\def\jref#1 #2 #3 #4 {{\par\noindent \hangindent=3em \hangafter=1
      \advance \rightskip by 0em #1, {\it#2}, {\bf#3}, #4.\par}}
\def\rref#1{{\par\noindent \hangindent=3em \hangafter=1
      \advance \rightskip by 0em #1.\par}}

\def\ch{ }

\usepackage{amssymb}

\begin{document}





 \title{The AGILE Mission}

             \author{
M. Tavani$^{1,2,3}$, G.~Barbiellini$^{4,5,3}$, A.~Argan$^{1}$,
F.~Boffelli$^{13}$, A.~Bulgarelli$^{8}$, P.~Caraveo$^{6}$,
P.~W.~Cattaneo$^{13}$, A.~W.~Chen$^{3,6}$, V.~Cocco$^{2}$,
E.~Costa$^{1}$, F.~D'Ammando$^{1,2}$, E.~Del~Monte$^{1}$,
G.~De~Paris$^{1}$, G.~Di~Cocco$^{8}$, G.~Di~Persio$^{1}$,
I.~Donnarumma$^{1}$, Y.~Evangelista$^{1}$, M.~Feroci$^{1}$,
A.~Ferrari$^{3,16}$, M.~Fiorini$^{6}$, F.~Fornari$^{6}$,
F.~Fuschino$^{8}$, T.~Froysland$^{3,7}$, M.~Frutti$^{1}$,
M.~Galli$^{9}$, F.~Gianotti$^{8}$, A.~Giuliani$^{3,6}$,
C.~Labanti$^{8}$, I.~Lapshov$^{1,15}$, F.~Lazzarotto$^{1}$,
F.~Liello$^{5}$, P.~Lipari$^{10,11}$, F.~Longo$^{4,5}$,
E.~Mattaini$^{6}$, M.~Marisaldi$^{8}$, M.~Mastropietro$^{24}$,
A.~Mauri$^{8}$, F.~Mauri$^{13}$, S.~Mereghetti$^{6}$,
E.~Morelli$^{8}$, A.~Morselli$^{7}$, L.~Pacciani$^{1}$,
A.~Pellizzoni$^{6}$, F.~Perotti$^{6}$, G.~Piano$^{1,2}$,
P.~Picozza$^{2,7}$, C.~Pontoni$^{3,5}$, G.~Porrovecchio$^{1}$,
M.~Prest$^{5}$, G.~Pucella$^{1}$, M.~Rapisarda$^{12}$,
A.~Rappoldi$^{13}$, E.~Rossi$^{8}$, A.~Rubini$^{1}$,
P.~Soffitta$^{1}$, A.~Traci$^{8}$, M.~Trifoglio$^{8}$,
A.~Trois$^{1}$, E.~Vallazza$^{5}$,
S.~Vercellone$^{6}$, V.~Vittorini$^{1,3}$, A.~Zambra$^{3,6}$, D.~Zanello$^{10,11}$\\
P. Giommi$^{14}$, S. Colafrancesco$^{14}$, C. Pittori$^{14}$,
B.~Preger$^{14}$, P. Santolamazza$^{14}$, F. Verrecchia$^{14}$, A. Antonelli$^{17}$\\
F. Viola$^{18}$, G. Guarrera$^{18}$, L.~Salotti$^{18}$,
F.~D'Amico$^{18}$, E.~Marchetti$^{18}$, M.~Crisconio$^{18}$\\
P.~Sabatini$^{19}$, G.~Annoni$^{19}$, S.~Alia$^{19}$,
A.~Longoni$^{19}$, R.~Sanquerin$^{19}$, M.~Battilana$^{19}$,
P.~Concari$^{19}$, E.~Dessimone$^{19}$, R.~Grossi$^{19}$,
A.~Parise$^{19}$
F.~Monzani$^{20}$, E.~Artina$^{20}$, R.~Pavesi$^{20}$,
G.~Marseguerra$^{20}$, L.~Nicolini$^{20}$, L.~Scandelli$^{20}$,
L.~Soli$^{20}$, V.~Vettorello$^{20}$, E.~Zardetto$^{20}$,
A.~Bonati$^{20}$, L. Maltecca$^{20}$, E.~D'Alba$^{20}$,
G.~Babini$^{21}$, F.~Onorati$^{21}$, L.~Acquaroli$^{21}$,
M.~Angelucci$^{21}$, B.~Morelli$^{21}$, C.~Agostara$^{21}$
M.~Cerone$^{22}$, A.~Michetti$^{22}$, P.~Tempesta$^{22}$,
S.~D'Eramo$^{22}$, F.~Rocca $^{22}$, F.~Giannini$^{22}$
G. Borghi$^{23}$, B.~Garavelli$^{25}$, M.~Conte$^{20}$,
M.~Balasini$^{20}$, I.~Ferrario$^{25}$, M.~Vanotti$^{25}$,
E.~Collavo$^{25}$, M.~Giacomazzo$^{25}$, M.~Patan\'e$^{26}$.
 }

\institute{
$^1$ INAF-IASF Roma, via del Fosso del Cavaliere 100, I-00133 Roma, Italy\\
$^2$ Dipartimento di Fisica, Universit\'a Tor Vergata, via della Ricerca Scientifica 1,I-00133 Roma, Italy\\
$^3$ Consorzio Interuniversitario Fisica Spaziale (CIFS), villa Gualino - v.le Settimio Severo 63,
I-10133 Torino, Italy\\
$^4$ Dip. Fisica, Universit\`a di Trieste, via A. Valerio 2, I-34127 Trieste, Italy\\
$^5$ INFN Trieste, Padriciano 99, I-34012 Trieste, Italy\\
$^6$ INAF-IASF Milano, via E. Bassini 15, I-20133 Milano, Italy\\
$^7$ INFN Roma Tor Vergata, via della Ricerca Scientifica 1, I-00133 Roma, Italy\\
$^8$ INAF-IASF Bologna, via Gobetti 101, I-40129 Bologna, Italy\\
$^9$ ENEA Bologna, via don Fiammelli 2, I-40128 Bologna, Italy\\
$^{10}$ INFN Roma 1, p.le Aldo Moro 2, I-00185 Roma, Italy\\
$^{11}$ Dip. Fisica, Universit\`a La Sapienza, p.le Aldo Moro 2, I-00185 Roma, Italy\\
$^{12}$ ENEA Frascati, via Enrico Fermi 45, I-00044 Frascati(RM),
Italy\\
$^{13}$ INFN Pavia, via Bassi 6, I-27100 Pavia, Italy\\
$^{14}$  ASI Science Data Center, ESRIN, I-00044 Frascati(RM),
Italy\\
$^{15}$ IKI, Moscow, Russia\\
$^{16}$ Dipartimento di Fisica, Universit\'a di Torino, Torino,
Italy\\
$^{17}$ Osservatorio Astronomico di Roma,  Monte Porzio Catone,
Italy
\\ $^{18}$  Agenzia Spaziale Italiana, viale Liegi , Rome, Italy\\
$^{19}$ Carlo Gavazzi Space, via Gallarate 139, 20151 Milano,
Italy,\\
$^{20}$ Thales Alenia Space (formerly Laben), S.S. Padana
Superiore 290, 20090
Vimodrone, Milano, Italy\\
$^{21}$  Rheinmetall Italia S.p.A. B.U. Spazio -
Contraves, via Affile, 102 00131 Roma,  Italy\\
$^{22}$ Telespazio, via Tiburtina 965, Italy\\
$^{23}$ Media Lario Technologies, Pascolo, 23842, Bosisio Parini
(Lecco), Italy\\
$^{24}$ CNR, IMIP, Montelibretti (Rome), Italy\\
$^{25}$ formerly at Carlo Gavazzi Space.\\
$^{26}$ formerly at Laben, S.S. Padana Superiore 290, 20090
Vimodrone, Milano, Italy}




 \abstract
 { AGILE is an Italian Space Agency mission dedicated to the
observation of the gamma-ray Universe. The AGILE very innovative
instrumentation combines for the first time a gamma-ray imager
(sensitive in the energy range 30 MeV - 50 GeV), a hard X-ray
imager (sensitive in the range 18-60 keV) together with a
Calorimeter (sensitive in the range 300 keV - 100 MeV) and an
anticoincidence system. AGILE was successfully launched on April
23, 2007 from the Indian base of Sriharikota and was inserted in
an equatorial orbit with very low particle background.}
 {AGILE  provides
crucial data for the study of Active Galactic Nuclei, GRBs,
pulsars, unidentified gamma-ray sources, Galactic compact objects,
supernova remnants,  TeV sources, and fundamental physics by
microsecond timing.}
   {An optimal angular resolution (reaching 0.1-0.2 degrees in
   gamma-rays and 1-2 arcminutes in hard X-rays)
and very large fields of view (2.5 sr and 1 sr, respectively) are
obtained by the use of Silicon detectors integrated in a very
compact instrument.}
   {AGILE surveyed the gamma-ray sky and detected many Galactic and extragalactic sources
   during the
   first months of observations. Particular emphasis is given to multifrequency observation programs of
   extragalactic and galactic objects.}
   { AGILE is a successful high-energy gamma-ray mission which reached its
   nominal scientific performance.
   The AGILE Cycle-1 pointing program started on 2007
December 1, and is open to the international community through a
Guest Observer Program. }

   \keywords{space missions --
                gamma-ray --
                X-ray
               }
\authorrunning{M. Tavani et al.}
\titlerunning{The AGILE Mission}

   \maketitle
%

\section{Introduction}
\label{}

Gamma-ray astrophysics above 100 MeV is an exciting field of
astronomical sciences  that received a strong impulse in recent
years. Detecting cosmic gamma-ray emission in the energy range
from a few tens of MeV to a few tens of GeV is possible only from
space instrumentation, and in the past 20 years several space
missions confronted the challenge of detecting  cosmic gamma-rays.
Gamma-ray emission from cosmic sources at these energies is
intrinsically non-thermal, and provides a diagnostic of particle
acceleration and  radiation processes in extreme conditions.
gamma-ray sources with the most energetic phenomena or objects of
our Universe including: Galactic compact objects and related
sources (e.g., pulsars, SNRs, pulsar wind nebulae, binary systems
with accreting neutron stars and black holes), molecular clouds
shining in gamma-rays because of interaction with energetic cosmic
rays, extragalactic massive black holes residing in Active
Galactic Nuclei (AGN), exploding stars originating Gamma-Ray
Bursts (GRBs).

The history of gamma-ray astronomy is a
topic beyond the scope of this paper. We briefly summarize here
the pioneering efforts and experiments that were carried out over
the years. The first detection of cosmic energetic radiation of
energy well above the electron's rest mass started in the late
Sixties with the detection of gamma-rays above 50 MeV by the OSO-3
satellite \citep{oso-3}. The first gamma-ray telescope with an
angular resolution of a few degrees was launched on November,
1972, by the Malindi site in Kenya as the second Small Astronomy
Satellite (SAS-2). A clear concentration of gamma-ray emission
from the Galactic plane was firmly established together with a
handful of pointlike gamma-ray sources
\citep{fichtel-1975,bignami-1979}. On August 8, 1975 the European
mission COS-B was launched and, after 7 years of observations,
several tens of gamma-ray sources were studied and catalogued
\citep{mayer-1979,swanenburg-1981}.

The Compton Gamma-Ray Observatory (CGRO) (active between
1991-2000) substantially increased our knowledge of the gamma-ray
Universe and provided a wealth of data on a large variety of
sources as well as unsolved puzzles. In particular, CGRO hosted
the Energetic Gamma-Ray Experiment Telescope (EGRET) operating in
the energy range 30 MeV-30 GeV which carried out a complete sky
survey detecting hundreds of gamma-ray sources
\citep{fichtel-1997,hartman-1999,th0,th1,th2,th3}. This scientific
inheritance is the starting point for any high-energy astrophysics
mission.

The new generation of high-energy space missions has to address
some of the fundamental issues that were left open (or unresolved)
by EGRET. For a mission following CGRO, it  is important to make
substantial progress in the instrumentation and mission concept to
achieve the following goals: (1) improving the gamma-ray angular
resolution near 100 MeV by at least a factor of 2-3 compared to
EGRET; (2) obtaining the largest possible field of view (FOV) at
100 MeV reaching 2.5-3 sr; (3) drastically reducing the deadtime
for gamma-ray detection from the EGRET value of 100 ms; (4)
obtaining broadband spectral information possibly including a
simultaneous detection capability in the MeV and X-ray energy
ranges; (5) carrying out a rapid quicklook analysis of the
gamma-ray data and a fast dissemination of results and alerts; (6)
 stimulating efficient multifrequency programs.
The AGILE Mission is aimed at meeting all these goals with an
unprecedented configuration for space gamma-ray astrophysics: a
Small Scientific Mission with optimized on-board and
ground-segment resources.

 The space program AGILE ({\it Astro-rivelatore
Gamma a Immagini LEggero}) is a high-energy astrophysics Mission
supported by the Italian Space Agency (ASI) with scientific and
programmatic participation by INAF, INFN, CNR, ENEA and several
Italian universities
\citep{tavani1,tavani-a,barbiellini-a,tavani-texas,tavani-2}. The
main industrial contractors include Carlo Gavazzi Space,
Thales-Alenia-Space (formerly Laben), Rheinmetall Italia (formerly
Oerlikon-Contraves), Telespazio, Galileo Avionica, and Mipot.

The main scientific goal of the AGILE program is to provide a
powerful and cost-effective mission with excellent imaging
capability simultaneously in the 30 MeV-50 GeV and 18-60 keV
energy ranges with a large field of view that is unprecedented in
high-energy astrophysics space missions.

\begin{figure}[t!]
\begin{center}
\includegraphics [height=12cm]{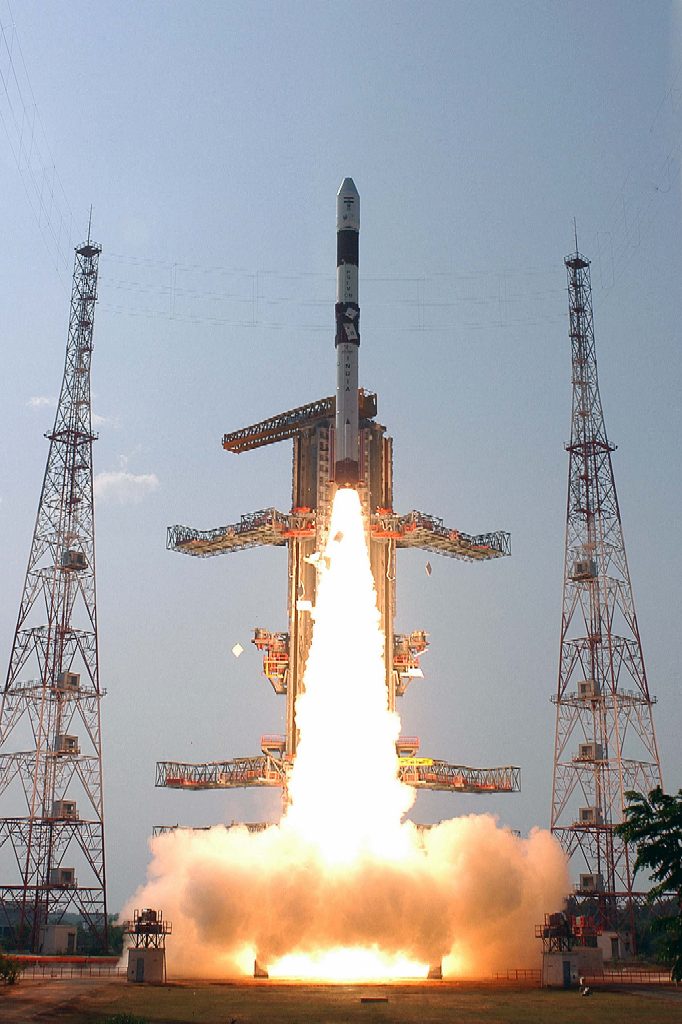} \caption { {\itshape The launch of the AGILE satellite by the Indian PSLV-C8
rocket from the Sriharikota base on April 23, 2007. } }
%
\end{center}
\label{lancio}
\end{figure}

AGILE was successfully launched by the Indian PSLV-C8 rocket from
the Sriharikota base on April 23, 2007 (see Fig.~\ref{lancio}).
The launch and orbital insertion were nominal, and a
quasi-equatorial orbit was obtained with the smallest inclination
(2.5 degrees) ever achieved by a high-energy space mission (e.g.,
SAS-2 and Beppo-SAX had inclinations of $\sim4$ degrees). The
satellite commissioning phase was carried out during the period
May-June, 2007. The scientific Verification phase and the in-orbit
calibration (based on long pointings at the Vela and Crab pulsars)
were carried out during the period July-November 2007. The nominal
scientific observation phase (AGILE Cycle-1, AO-1) started on
December 1, 2007. AGILE is the first gamma-ray Mission
successfully operating in space after the long temporal hiatus of
almost ten years since the end of EGRET operations. It will
operate together with the GLAST mission that was launched on
June~11,~2008 \citep{glast}.



\begin{figure}[t!]
\begin{center}
\includegraphics [height=9cm]{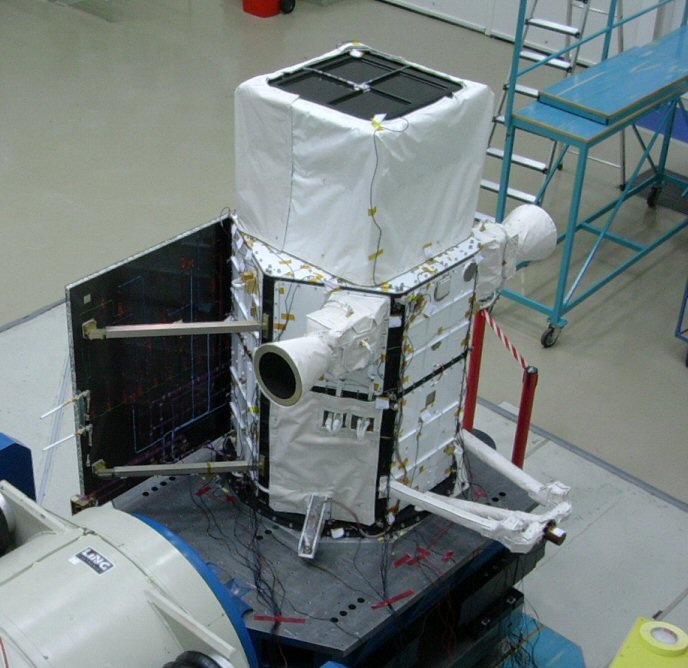} \caption { {\itshape
    The integrated AGILE satellite in its final
configuration being covered by the thermal blanket during the
qualification tests in IABG (Munich), July 2006. The total
satellite mass is equal to 350 kg. } \label{sat-1}}
\end{center}
\end{figure}

The AGILE instrument design is very innovative and based on
 solid state Silicon detector technology and state-of-the-art
 electronics  and readout systems developed in Italian laboratories
\citep{barbiellini-1,barbiellini-2,nina}.
The  instrument is light ($\sim 100$~kg) and very compact (see
Fig.~\ref{PL-1}).
The total satellite mass is about 350 kg (see Fig.~\ref{sat-1}).

AGILE is expected to substantially advance our knowledge in
several research areas including the study of Active Galactic
Nuclei and massive black holes, Gamma-Ray Bursts (GRBs), the
unidentified gamma-ray sources, Galactic transient and steady
compact objects, isolated and binary pulsars, pulsar wind nebulae
(PWNae), supernova remnants, TeV sources, and the Galactic Center
while mapping the overall gamma-ray emission from our Galaxy.
Furthermore, the fast AGILE electronic readout and data processing
(resulting in detector deadtimes smaller than $\sim $ 200
$\mu$sec) allow to perform, for the first time, a systematic
search for sub-millisecond gamma-ray/hard X-ray transients that
are of interest for both Galactic compact objects (searching
outburst durations comparable with the dynamical timescale of
$\sim 1 \, M_{\odot}$ compact objects) and quantum gravity studies
in extragalactic sources.

The AGILE Science Program is aimed at providing a complete sky
coverage during its first year while allowing a
 prompt response to gamma-ray transients and alert for follow-up
multiwavelength observations.
 AGILE provides important  information  complementary to several
high-energy space missions (Chandra, INTEGRAL, RXTE,  XMM-Newton,
SWIFT, {\ch Suzaku}, GLAST) and  supports ground-based
investigations in the radio, optical, and TeV bands. Part of the
AGILE Science Program is open for Guest Investigations on a
competitive basis. Quicklook data analysis and fast communication
of new transients has been implemented as an essential part of the
AGILE Science Program.

\section{Mission Concept}

The AGILE program is motivated by very specific scientific
requirements and goals.
The essential point, that permeates the whole mission from its
conception, is to provide a very effective gamma-ray space
instrument with excellent detection and imaging capabilities both
in the gamma-ray and hard X-ray energy ranges. The very stringent
mission constraints (satellite and instrument volume, weight,
cost, and optimized ground segment) determined, from the very
beginning of the mission development, a specific optimization
strategy.

The AGILE mission challenge has been obtaining optimal
gamma-ray/hard X-ray detection capabilitity (together with
excellent timing resolution in the energy band near 1 MeV) with a
very light ($\sim 100$~kg) instrument.

The AGILE instrument has been therefore designed and developed to
obtain:

\begin{itemize}
\item {\bf excellent imaging capability in the energy range 100~MeV-50~GeV},
improving the EGRET angular resolution by a factor of 2;

\item {\bf a very large field-of-view} for both the gamma-ray imager (2.5 sr, i.e.,
$\sim$5 times larger than that of EGRET) and the hard X-ray imager
(1 sr);

\item {\bf excellent  timing capability}, with overall photon absolute time
tagging uncertainty of $2 \, \mu$s  coupled with very small
deadtimes for gamma-ray detections;

\item {\bf a good sensitivity for pointlike gamma-ray and hard X-ray sources}.
Depending on exposure and background, after a 1-year program the
flux sensitivity threshold can reach values of $(10-20)\times
10^{-8} \; \rm photons \, \rm cm^{-2} \, s^{-1}$ at energies above
100 MeV. The hard X-ray imager sensitivity is between 15 and 50
mCrab at 20 keV for a 1-day exposure over a 1~sr field of view;

\item {\bf good sensitivity to photons in the energy range
$\sim$30-100~MeV}, with an effective area above $200 \; cm^2$ at
30~MeV;

\item {\bf a  rapid response to gamma-ray transients and
gamma-ray bursts}, obtained by a special quicklook analysis
program and coordinated ground-based and space observations;

\item {\bf accurate localization ($\sim$2-3 arcmins) of
GRBs and other transient events} obtained by the GRID-SA
combination (for typical hard X-ray transient fluxes above
$\sim$1~Crab); the expected GRB detection rate for AGILE is $\sim
1-2$ per month;

\item {\bf long-timescale continuous monitoring ($\sim$2-3 weeks) of
gamma-ray and hard X-ray sources};

\item {\bf satellite repointing within $\sim 1$ day after special
alerts}.

\end{itemize}

The simultaneous hard X-ray and gamma-ray observations represent a
novel approach to the study of high-energy sources. In the
following, we address the  main relevant features of the AGILE
scientific performance.

\section{Scientific Objectives}

AGILE fits into the discovery path followed by previous gamma-ray
missions (SAS-2, COS-B, and EGRET) and is complementary to GLAST.
Nearly 270 gamma-ray sources above 30~MeV  were
catalogued\footnote{See also the analysis of Casandjian \&
Grenier, 2008.} by EGRET (with only 30\% identified as AGNs or
isolated pulsars) \citep{hartman-1999}.
We summarize here the main AGILE's scientific objectives.

$\bullet$ {\bf Active Galactic Nuclei}. For the first time, it is
possible to monitor tens of potentially gamma-ray emitting
 AGNs during each pointing. Several outstanding issues
concerning the mechanism of AGN gamma-ray production  and time
evolution can be addressed by AGILE
 including:
(1) the study of transient vs. low-level gamma-ray emission and
duty-cycles \citep{vercellone}; (2) the relationship between the
gamma-ray variability and the radio-optical-X-ray-TeV emission;
(3) the possible correlation between relativistic radio plasmoid
ejections and gamma-ray flares; (4) hard X-ray/gamma-ray
correlations.

$\bullet$ {\bf Gamma-ray bursts}.
 A few  GRBs were
detected by the  EGRET spark chamber \citep{Schneid96a}.
This number appeared to be limited by the EGRET FOV and
sensitivity and not by the intrinsic GRB emission mechanism. Owing
to a larger FOV, the GRB detection rate by the AGILE-GRID is
expected to be larger than that of EGRET, i.e., $\sim$2--5
events/year. Furthermore,
 the small GRID deadtime (${\ch \sim 500 }$  times smaller than
that of EGRET) allows for a  better study of the initial phase of
GRB pulses (for which EGRET response was in some cases
inadequate).
The hard X-ray imager (Super-AGILE) can localize GRBs within a few
arcminutes, and can systematically study the interplay between
hard X-ray
 and gamma-ray emissions. Special emphasis is given
to the {\ch search for sub-millisecond  GRB pulses independently
detectable by the Si-Tracker, {\ch MCAL} and Super-AGILE.}

$\bullet$ {\bf Diffuse Galactic  gamma-ray emission}. The AGILE
good angular resolution and large average exposure  further
improves our knowledge of cosmic ray origin, propagation,
interaction and emission processes. A detailed gamma-ray imaging
of individual molecular cloud complexes is possible.

$\bullet$ {\bf Gamma-ray pulsars and PWNae}. AGILE  will
contribute to the study of gamma-ray pulsars (PSRs) in several
ways: (1) improving timing and lightcurves of known gamma-ray
PSRs; (2) improving photon statistics for blind gamma-ray period
searches of pulsar candidates;  (3) studying unpulsed gamma-ray
emission from plerions in supernova remnants and studying  pulsar
wind/nebula interactions, e.g., as in the Galactic sources
recently discovered in the TeV range \citep{aharonian}.
Particularly interesting for AGILE are the $\sim30$ new young PSRs
discovered in the Galactic plane by the Parkes survey
\citep{kramer}.

$\bullet$ {\bf Search for non-blazar gamma-ray variable sources in
the Galactic plane}, currently a new class of unidentified
gamma-ray sources such as GRO~J1838-04 \citep{tavani0}.

$\bullet$ {\bf Compact Galactic sources, micro-quasars, new
transients}. A large number of gamma-ray sources near  the
Galactic plane are unidentified, and sources such as
2CG~135+1/LS~I+61~303 can be monitored on timescales of months.
Cyg X-1 is also monitored and gamma-ray emission above 30 MeV will
be intensively searched. Galactic X-ray jet sources (such as Cyg
X-3, GRS~1915+105, GRO~J1655-40 and others) can produce detectable
gamma-ray emission for favorable jet geometries, and a TOO program
is planned to follow-up new discoveries of {\it micro-quasars}.

$\bullet$ {\bf Supernova Remnants (SNRs)}. Several possible
gamma-ray source-SNR associations were proposed based on EGRET
data \citep{sturner,esposito}. However, none are decisive.
High-resolution imaging of SNRs in the gamma-ray range can provide
the missing information to decide between leptonic and hadronic
models of SNR emission above 100 MeV.


$\bullet$ {\bf Fundamental Physics: Quantum Gravity}. AGILE
detectors are suited for Quantum Gravity (QG) studies.
The existence of sub-millisecond GRB pulses lasting hundreds of
microseconds \citep{bhat} opens the way to study QG delay
propagation effects by AGILE detectors. Particularly important is
the AGILE Mini-Calorimeter with photon-by-photon independent
readout for each of the 30 CsI bars of small deadtime ($\sim 20 \,
\mu$s) and absolute timing resolution
 ($\sim 3 \, \mu$s). Energy dependent
time delays near $\sim 100 \, \mu$s for ultra-short GRB pulses in
the energy range 0.3--3 MeV can be detected.
If these GRB ultra-short pulses originate at cosmological
distances, sensitivity to the Planck's mass can be reached by
AGILE.

\section{Main Characteristics of the Mission}

In order to make substantial progress with respect to the previous
generation of gamma-ray astrophysics missions and to substantially
contribute to the current and future observations, the AGILE
instrument is required to achieve an optimal performance with the
following characteristics.

\subsection{Gamma-ray and X-ray angular  resolution}

The gamma-ray detection resolution
 is required to achieve an effective PSF with 68\% containment radius
of $ 1^{\circ}-2^{\circ}$ at E $>$ 300~MeV allowing a gamma-ray
source positioning with error box radius near
 6$^{'}-20^{'}$ depending on source spectrum, intensity, and
sky position. The hard X-ray imager operating in the 18-60~keV
band has a spatial resolution of 3 arcminute (pixel size). This
translates into a pointing reconstruction of 1-2 arcmins for
relatively strong transients at the Crab flux level.

\subsection{Large FOVs of the gamma-ray and hard X-ray imagers}

A crucial feature of the AGILE instrument is its very large field
of view for both the gamma-ray and hard X-ray detectors. The
gamma-ray FOV is required to be 2.5-3 sr, i.e., to cover about 1/5
of the entire sky. The hard X-ray imager is required to cover a
region of about 1 sr. The combination of coaligned gamma-ray and
hard X-ray imagers with very large FOVs is unprecedented and it is
the main novelty of the AGILE instrument configuration.

\subsection{Fast reaction to strong high-energy transients}

The existence of a large number of variable gamma-ray sources,
e.g., near the Galactic plane, \cite{tavani0}, requires  a
reliable program for quick response to transient gamma-ray
emission. Quicklook Analysis of gamma-ray data is an important
task of the AGILE Mission.
 Prompt communication of bright
gamma-ray transients (above $10^{-6} \rm \, ph \, cm^{-2} \,
s^{-1}$, requiring typically 1-2 days to be detected with high
confidence) is  ensured by a proper Ground Segment configuration.
Alerts for  shorter timescale (seconds/minutes/hours) transients
(GRBs, SGRs, and other bursting events) are also possible. A
primary responsibility of the AGILE Team is to provide positioning
of short-timescale transient as accurate as possible, and to alert
the community though dedicated channels.

\subsection{A large exposure for Galactic and extragalactic
sources}

Owing to a larger effective area at large off-axis angles, the
AGILE average exposure above 100 MeV will be typically 4 times
larger than that of EGRET after a 1-year sky survey.
 Deep exposures for
selected regions of the sky can be obtained  with repeated
overlapping pointings.  This can be particularly useful to monitor
simultaneously selected  Galactic and extragalactic sources.

\subsection{High-Precision Timing} AGILE detectors have optimal
timing capabilities. The on-board navigation system makes possible
an absolute time tagging precision for individual photons of $2 \,
\mu $s. Depending on the event characteristics, absolute time
tagging can achieve values near $2 \, \mu$s for the
Silicon-Tracker, and $3-4 \, \mu$s for the Mini-Calorimeter and
Super-AGILE.

The instrumental AGILE deadtime is unprecedentedly small for
gamma-ray detection (typically less than $200 \, \mu$s).
 Taking into account the segmentation of the electronic
readout of {\ch MCAL} and Super-AGILE detectors (30 {\ch MCAL}
elements and 16 Super-AGILE elements) the {\ch  MCAL and SA }
effective deadtimes are substantially less than for the individual
units. {\ch We  reach $\sim 2 \, \mu$s for MCAL, and $5 \, \mu$s
for SA.} Furthermore, a special memory  ensures that MCAL events
detected during the Si-Tracker readout deadtime are automatically
stored in the GRID event. For these events, precise timing and
detection in the $\sim 1$--200~MeV range can be achieved with
temporal resolution well below $ 100  \, \mu$s. This may be of
great relevance for AGILE high-precision timing investigations.

\section{The Scientific Instrument}

\begin{figure}  
\begin{center}
\includegraphics [height=7.5cm]{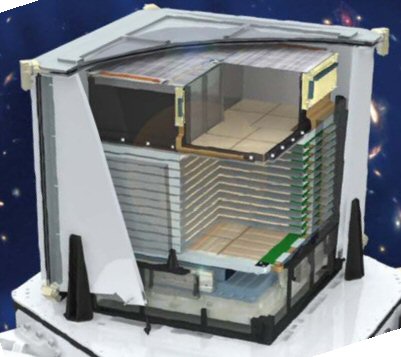} \caption { {\itshape
    The AGILE scientific instrument showing the hard X-ray
imager, the gamma-ray Tracker, and Calorimeter. The
Anticoincidence system is partially displayed, and no lateral
electronic boards and harness are shown for simplicity. The AGILE
instrument "core" is approximately a cube of about 60 cm size and
of weight approximately equal to 100 kg. } \label{PL-1}}
\end{center}
\end{figure}

The AGILE scientific payload is made of three detectors surrounded
by an Anticoincidence system, all combined into one integrated
instrument with broad-band detection and imaging capabilities. A
dedicated Data Handling system completes the instrument.
Fig.~\ref{PL-1} shows a schematic view of the instrument, and
Table~1 summarizes the main instrument scientific performance. We
summarize here the main characteristics of the instrument: several
papers describe the individual detectors in detail
\citep{perotti,barbiellini-1,Prest_2003,feroci,labanti_2006}.

 {\bf The  Gamma-Ray Imaging Detector (GRID)} is sensitive in
the energy range $\sim 30$~MeV--$50$~GeV, and consists of a
Silicon-Tungsten Tracker, a Cesium Iodide Calorimeter, and an
Anticoincidence system.

The GRID trigger logic and data acquisition system (based on
Anticoincidence, Tracker and Mini-Calorimeter information) allows
for an efficient background discrimination and inclined photon
acceptance \citep{tavani-dh,argan-dh}. The GRID is designed to
achieve an optimal angular resolution (source location accuracy
$\sim6'-12'$ for intense sources), a very large field-of-view
($\sim 2.5$~sr), and a sensitivity comparable  to that of EGRET
for sources within 10-20 degree from the main axis direction (and
substantially better for larger off-axis angles).

{\bf The hard X-ray Imager (Super-AGILE)} is a unique feature of
the AGILE instrument (for a complete description, see
\cite{feroci}).
 The imager is placed on
top of the gamma-ray detector and is sensitive in the 18-60 keV
band.

{\ch {\bf A Mini-Calorimeter operating in the "burst mode"} is the
third AGILE detector.} It is part of the GRID, but also also
capable of independently detecting GRBs and other transients in
the 350 keV - 50 MeV energy range with excellent timing
capabilities.

 \begin{table*}    
 \begin{center}
 \large

 \centerline{\bf Table 1 -  The AGILE Scientific Performance } \vskip .08in
 \label{table-1}
 \begin{tabular}{|lc|}
 \hline
   {\bf Gamma-ray Imaging Detector (GRID)} &      \\
 \hline
 Energy range & 30 MeV -- 50 GeV    \\
 Field of view &  $\sim 2.5$  sr    \\
 Flux sensitivity ($E > 100$ MeV, 5$\sigma$ in 10$^{6}$ s)
  &  3$\times$10$^{-7}$  (ph
 cm$^{-2}$ s$^{-1}$)   \\
 Angular resolution  at 100 MeV  (68\% cont. radius)   &  3.5 degrees   \\
 Angular resolution  at 400 MeV  (68\% cont. radius)   &  1.2 degrees   \\
 Source location accuracy (high Gal. lat., 90\% C.L.))  &  $\sim$15 arcmin  \\
 Energy resolution   (at 400 MeV)      & $\Delta$E/E$\sim$1   \\
 Absolute time resolution    & $\sim 2 \, \mu$s  \\
 Deadtime            & $\sim  100-200 \, \mu$s \\
 \hline
  {\bf  Hard X--ray Imaging Detector (Super-AGILE)}  &      \\
 \hline
 Energy range & 18 -- 60 keV   \\
 Single (1-dim.) detector FOV (FW at zero sens.) &  107$^{\circ}\times$68$^{\circ}$   \\
 Combined (2-dim.) detector FOV  (FW at zero sens.)&  68$^{\circ}\times$68$^{\circ}$   \\
 Sensitivity (18-60 keV, 5$\sigma$ in 1 day)   &  $\sim${\ch 15} mCrab    \\
 Angular resolution (pixel size)     &   6 arcmin    \\
 Source location accuracy (S/N$\sim$10)  &  $\sim$1-2 arcmin     \\
 Energy resolution {\ch(FWHM)}          & $\Delta$E $\sim$ {\ch 8} keV     \\
 Absolute time resolution & $\sim 2 \, \mu$s \\
 \hline
 {\bf Mini-Calorimeter}  &  \\
 \hline
 Energy range        & 0.35 -- 50  MeV   \\
 Energy resolution ({\ch at 1.3 MeV })  &  {\ch 13\% FWHM }   \\
  Absolute time resolution    & $\sim 3 \, \mu$s  \\
 Deadtime (for each of the 30 CsI bars)    & $\sim 20 \, \mu$s \\
 \hline
 \end{tabular}
 \end{center}
 \label{tab-payload}
 \end{table*}

Fig.~\ref{sat-1} shows  the integrated AGILE satellite and
Fig.~\ref{PL-1} a schematic representation of the instrument. We
briefly describe here the main detecting units of the AGILE
instrument; more detailed information will be presented elsewhere.

\subsection{\bf The Anticoincidence  System}

The Anticoincidence (AC) System is  aimed at a very efficient
charged particle background rejection \citep{perotti}; it also
allows a preliminary direction reconstruction for triggered photon
events through the DH logic. The AC system completely surrounds
all AGILE detectors (Super-AGILE, Si-Tracker and {\ch MCAL}). Each
lateral face
 is segmented in three plastic scintillator
layers ($0.6$~cm thick) connected to photomultipliers placed at
the bottom of the panels.
 A single plastic scintillator layer
(0.5~cm thick) constitutes the top-AC whose signal is read by four
light photomultipliers  placed at the four corners of the
structure frame. The segmentation of the AC System and the  ST
trigger logic
 contribute in an essential way to produce  the very large field
of view of the AGILE-GRID.

\subsection{\bf The Silicon-Tracker}

\begin{figure}[t!]
\begin{center}
\vspace*{-0.2cm} \hspace*{-2.5cm} \includegraphics
[height=10.5cm]{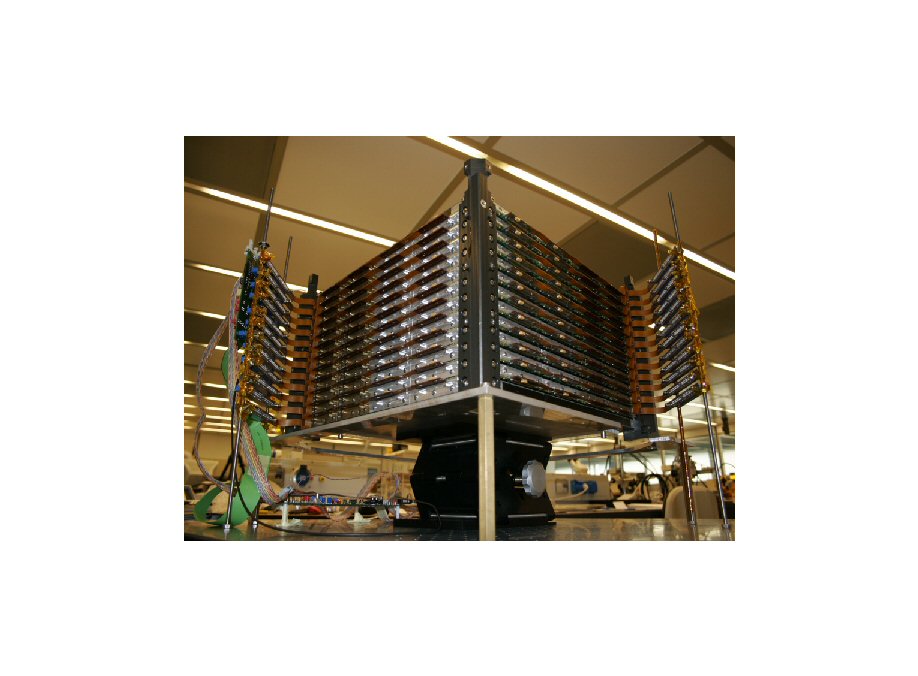} \vspace*{-2cm} \caption {
{\itshape
 The assembled AGILE Silicon Tracker developed in the
Trieste INFN laboratories before being integrated with the rest of
the instrument (June 2005).     } \label{agile-PL-4}}
\end{center}
 \end{figure}

\begin{figure}  [h!]
\begin{center}
\includegraphics [height=13.5cm]{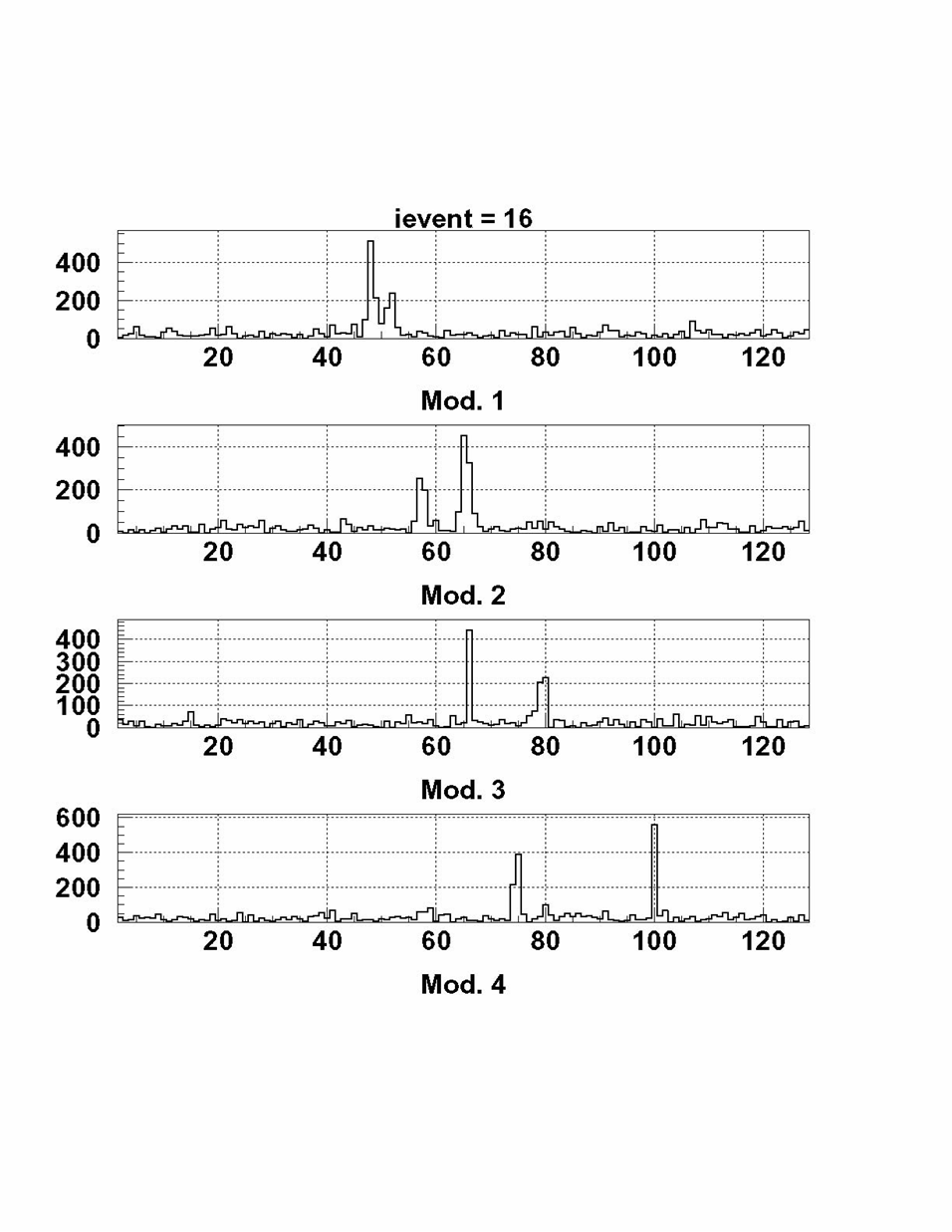}
\vspace*{-2.cm} \caption { {\itshape Detection of a typical
gamma-ray event by the AGILE Tracker in the Thales-Alenia Space
laboratory in Milan during integration tests (July 2005). The
electron-positron hits in the Silicon microstrips produce typical
"clusters" of read-out strips whose analogue signals produce a
sequence of deposited energy "histograms". This cluster
positioning capability is a unique feature of the AGILE gamma-ray
imager. } \label{agile-PL4b}}
\end{center}
 \end{figure}

The Silicon Tracker  (ST) is   the AGILE gamma-ray imager  based
on photon conversion into electron-positron pairs
\citep{barbiellini-2,Prest_2003}. It consists of a total of 12
trays with a repetition pattern of {\ch 1.9}~cm
(Fig.~\ref{agile-PL-4}). The first 10 trays are capable of
converting gamma-rays by a Tungsten layer. Tracking of charged
particles is ensured by Silicon microstrip detectors that are
configured to provide the two orthogonal coordinates for each
element (point) along the track. The individual Silicon detector
element is a tile of area
 ${ {9.5}\times{9.5} }$ cm$^{2}$, microstrip pitch of 121~$\mu$m,
 and  410~$\mu$m thickness. Four Silicon tiles are bonded together
 to provide a ladder. Four ladders constitute a ST plane.
The AGILE ST readout system is capable of detecting and storing
the energy deposited in the Silicon microstrips by the penetrating
particles. The readout signal is processed for half of the
microstrips by an alternating readout system characterized by
"readout" and "floating" strips. The analog signal produced in the
readout strips is read and stored for further processing. Each
Silicon ladder has a total of 384 readout channels (242~$\mu$m
readout pitch) and 3 TAA1 chips are required to process
independently the analog signal from the readout strips. Each
Si-Tracker layer is then  made of $4 \times 4$ Si-tiles, for a
total geometric area of ${ {38}\times{38}}$ cm$^{2}$. The first 10
trays are equipped with a Tungsten layer of 245 $\mu$m ($0.07 \;
X_0$) positioned in the bottom part of the tray.  The two
orthogonal coordinates of particle hits in the ST are provided by
two layers of Silicon detectors properly configured for each tray
that therefore has  $2 \times 1,536$ {readout microstrips}. Since
the ST trigger requires a signal from at least three (contiguous)
planes, two more trays are inserted at the bottom of the Tracker
without the Tungsten layers. The total readout channel number for
the GRID Tracker is then ${36,864}$.
The 1.9~cm distance between mid-planes has been optimized through
extensive  Montecarlo simulations. The ST has an {\it on-axis}
total radiation length near $0.8 \; X_0$.
Special trigger logic algorithms implemented on-board (Level-1
and Level-2) lead to a substantial particle/albedo-photon
background subtraction and a preliminary on-board reconstruction
of the photon incidence angle. Both digital and analog information
are crucial for this task.
Fig.~\ref{agile-PL4b} shows a typical read-out configuration of a
gamma-ray event detected by the AGILE Silicon Tracker. The
positional resolution obtained by the ST is excellent, being below
40 $\mu$m for a wide range of particle incidence angles
\cite{barbiellini-5}.

\subsection{\bf Super-AGILE}

\begin{figure}[t!]
\begin{center}
\includegraphics [height=6.5cm]{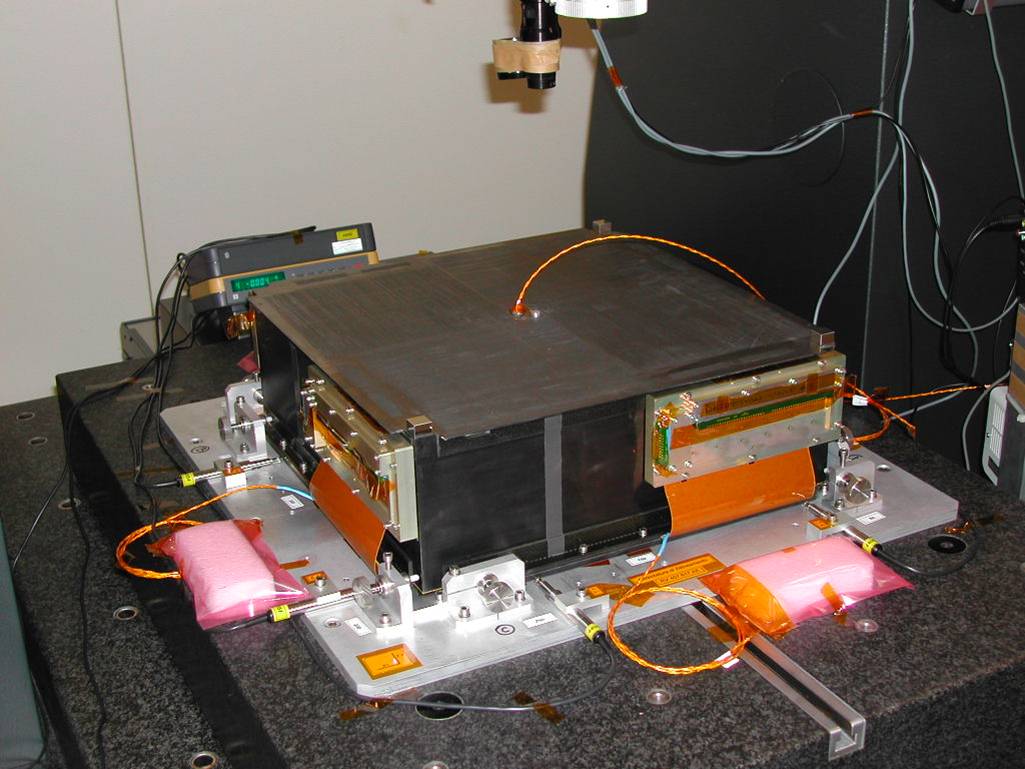} \caption { {\itshape
 The Super-AGILE detector in the INAF-IASF laboratory during metrology measurements
(March 2005).    }  \label{agilePL05}}
\end{center}
\end{figure}

Super-AGILE  (SA), the ultra-compact and light hard-X-ray imager
of AGILE \citep{feroci} is a coded-mask system made of a Silicon
detector plane and a thin Tungsten mask positioned 14 cm above it
(Fig.~\ref{agilePL05}). The detector plane is organized in four
independent square Silicon detectors ($ 19 \times 19 \, \rm cm^2$
each) plus dedicated front-end electronics based on the {\ch
XAA1.2} chips suitable to the SA energy range \cite{delmonte2}.
The total number of SA readout channels is { 6,144}. The detection
cabability of SA includes: (1) photon-by-photon {\ch transmission}
and imaging of sources in the energy range 18-60~keV, with a large
field-of-view (FOV $\sim 1$~sr); (2) an angular resolution of 6
arcmin; (3) a good sensitivity ($\sim 15${\ch ~mCrab between 18-60
keV} for 50 ksec integration, and $\loe 1$~Crab for a few seconds
integration). SA is aimed at the  hard X-ray detection
simultaneously with gamma-ray
 detection of high-energy sources with excellent timing capabilities
 (a few microseconds). The SA acquisition logic produces on-board essential
 GRB quantities such as time,  coordinates and preliminary flux estimates.
 The AGILE
 satellite is equipped with an ORBCOMM transponder capable of
 trasmitting the GRB on-board processed physical quantities
  to the ground within {\ch 10-30 min}.

\subsection{\bf The Mini-Calorimeter}

\begin{figure}[h!]
\begin{center}
\includegraphics [height=5.5cm]{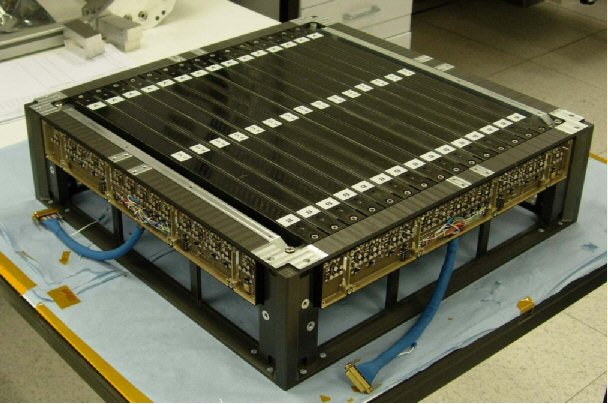} \caption { {\itshape
 The MCAL detector consisting of two planes of CsI bars enclosed in Carbon fiber supporting structure
 is shown during the integration process with the signal readout diodes and FFE in the Thales-Alenia Space laboratories
 in Milan (February, 2005).    }  \label{agilePL06}}
\end{center}
\end{figure}

The Mini-Calorimeter (MCAL) {\ch is made of 30  Caesium Iodide
(CsI(Tl)) } bars arranged in two planes, for a total (on-axis)
radiation length of $1.5 \; X_0$ (see Fig.~\ref{agilePL06}). A
detailed description of the MCAL detector can be found in
\cite{labanti_2006,labanti_2008}. The signal from each CsI bar is
collected by two photodiodes placed at both ends. The MCAL aims
are: {\it (i)} obtaining information on the energy deposited in
the CsI bars by particles produced in the Silicon Tracker (and
therefore contributing to the determination of the total photon
energy);
{\it (ii)} detecting GRBs and other impulsive events with spectral
and intensity information in the energy band $\sim 0.35 -
100$~MeV. An independent {\ch burst search algorithm is}
implemented on board with a wide range of trigger timescales for
an MCAL independent GRB detection. Following a GRB trigger, MCAL
is indeed able to store photon-by-photon information for a
duration dynamically determined by the on-board logic. The MCAL
segmentation and the photon-particle hit positioning along the
bars allow to obtain the general configuration of "hits" across
the calorimeter volume. This information is used by the on-board
trigger logic for background discrimination and by the ground
processing to obtain a preliminary determination of GRB direction.

\subsection{\bf Data Handling and Power Supply}

 The Data Handling (DH) and power supply systems complete the
 instrument. The DH is optimized
for fast on-board processing of the GRID, Mini-Calorimeter and
Super-AGILE data \citep{argan-dh2,tavani-dh,argan-dh}. Given the
relatively large number of readable channels in the ST and
Super-AGILE ($\sim$40,000), the instrument
 requires a very efficient on-board data processing system.
 The GRID trigger logic for the acquisition of
gamma-ray photon data and background rejection is structured in
two main levels: Level-1 and Level-2 trigger stages. The Level-1
trigger is fast ($\loe 5 \mu$s) and requires a signal in at least
three out of four contiguous tracker planes, and a proper
combination of fired TAA1 chip number  and AC signals. An
intermediate Level-1.5 stage is also envisioned (lasting $\sim
20~\mu$s), with the acquisition of the event topology based on the
identification of fired TAA1 chips. Both Level-1 and Level-1.5
have a hardware-oriented veto logic providing a first cut of
background events. Level-2 data processing includes a GRID readout
and pre-processing, ``cluster data acquisition" (analog and
digital information).
The Level-2 processing is asynchronous (estimated duration $\sim$
a few ms) with the actual GRID event processing. The GRID
deadtime  turns out to be $\loe 200 \; \mu$s and is dominated by
the Tracker readout.

The charged particle and albedo-photon background passing the
Level-1+1.5 trigger level of processing is measured in orbit to be
$\loe 100$~events/sec for the equatorial orbit of AGILE. The
on-board Level-2 processing  has the task of reducing
 this background  by a factor between 3 and 5.
Off-line ground processing of the GRID data has the goal to reduce
the particle and albedo-photon background rate above 100 MeV  to
the expected rate of $\sim $0.02~events/sec.

In order to maximize the GRID FOV and detection efficiency for
large-angle incident gamma-rays (and  minimize the effects of
particle backsplash from the MCAL and of ``Earth albedo"
background photons), the data acquisition logic  uses proper
combinations of top and lateral AC signals and a coarse  on-line
direction reconstruction in the ST. For laterally incident events
depositing more than 200 MeV in the {\ch MCAL}, the AC veto may be
disabled to allow the acquisition of gamma-ray photon events with
energies larger than 1~GeV.

A special set of memory buffers and burst search algorithms are
implemented to maximize data acquisition for transient gamma-ray
events (e.g., GRBs) in the ST, Super-AGILE and Mini-Calorimeter,
respectively. The Super-AGILE event acquisition envisions a first
``filtering" based on AC-veto signals, and pulse-height
discrimination in the front end electronics. The events are then
buffered and transmitted to the CPU for burst searching and final
data formatting. The four Si-detectors of Super-AGILE
 are organized in sixteen independent readout units,
 {\ch resulting in a $\sim 5 \; \mu$s global deadtime}.

In order to maximize the detecting area and minimize the
instrument  weight, the GRID and Super-AGILE front-end-electronics
 is partly accommodated in special boards placed externally on the
Tracker lateral sides.
 Electronic boxes, P/L memory (and buffer)
units are positioned at the bottom of the instrument within the
spacecraft.

\subsection{The AGILE instrument calibration}

The AGILE scientific instrument was fully calibrated on the ground
during a set of  calibration campaigns dedicated to the three
instrument detectors. We briefly summarize here the main
calibration operations, postponing a detailed analysis of the
campaigns to forthcoming papers.

\subsubsection{The gamma-ray imager calibration}

The AGILE-GRID calibration was carried out at the INFN National
Laboratories in Frascati during the period November 1-25, 2005.

A beam of gamma-ray photons in the energy range 20-700 MeV was
produced by
    Bremsstrahlung of electrons and tagged by a dedicated set-up in the
    Beam Test Facility  of the INFN Laboratori Nazionali di Frascati
    based on the measurement with silicon strip detectors of the electron
    trajectory in a magnetic field.
A total of 100,000 tagged events was accumulated for several
incidence directions and instrument configurations.
Both the GRID spectral and PSF response were carefully
studied and compared with results of extensive simulations
\citep{pucella,longo}. Furthermore, the leptonic background was
studied by using the direct electron and positron beams
interacting with the AGILE GRID for different geometries. A
sequence of runs was obtained for both direct incidence on the
instrument as well as for events originating by interactions with
the spacecraft.

\subsubsection{The hard X-ray imager calibration}

The Super-AGILE imager was calibrated at different stages during
the instrument integration and testing. It was first calibrated at
the detection plane and stand-alone detector level in the clean
room of INAF-IASF Rome on April and August 2005, respectively. The
SA effective area and intrinsic imaging properties were
investigated by means of a highly collimated X-ray tube and
point-like radioactive sources (see Donnarumma et al. 2006,
Evangelista et al. 2006 for details). A dedicated procedure was
developed (Donnarumma et al. 2008) to correct the SA images for
the beam divergence in order to derive imaging calibration
properties from measurements with radioactive sources at finite
distance (about 200 cm from the experiment).

\begin{figure}[h!]
\begin{center}
\includegraphics [height=9.5cm]{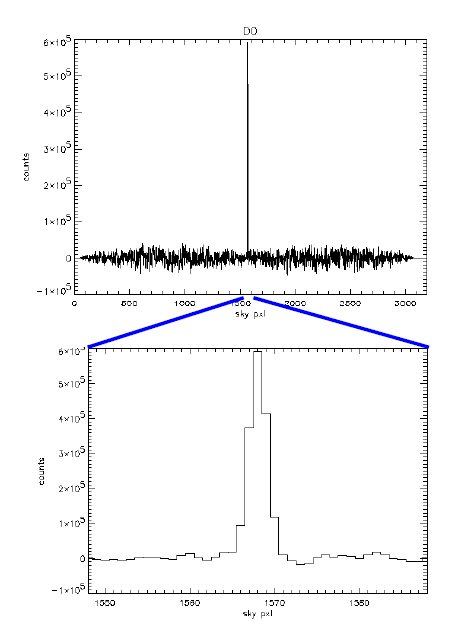} \caption { {\itshape
Calibration data of the Super-AGILE imaging detector obtained with
a 22 keV radioactive source placed at a 200 cm distance, after
correcting for beam divergence effects.
   }  \label{SA-calib}}
\end{center}
\end{figure}

Once integrated with the instrument and satellite, the SA imager
was then fully calibrated during in January, 2007 at the CGS
facility in Tortona. A sequence of measurements were carried out
with radioactive sources positioned at different angles with
respect to the instrument axis. The imaging response  was studied
as a function of the source position in the field of view (in more
than 40 positions), and energy (at 22, 30 and 60 keV). Radioactive
sources were held at 200 cm distance and their position were
independently measured by Super-AGILE and by an optical laser
tracking system (in collaboration with the metrology group of ENEA
Frascati). The calibration campaign envisaged a total of more than
110 measurements and 340 ksec livetime, with more than 10$^7$
source photons collected. The data analysis allowed us to
calibrate the imaging and spectral response of SA.
Fig.~\ref{SA-calib} shows a sample of the results achieved during
the final ground calibrations, confirming the expected 6 arcmin
(FWHM) point spread function (PSF) and the $\sim$1-2 arcmin point
source location accuracy (see Feroci et al. 2007 and Evangelista
et al. 2008 for more details). SA on-board imaging was also tested
in Tortona. A sequence of GRB simulating tests were carried out to
check the on-board trigger logic and parameter setting.

\subsubsection{The Calorimeter calibration}

 Several calibration sessions of MCAL were carried out after its integration
\citep{labanti_2008}. A first stand-alone calibration session was
performed at instrument level, prior to integration into the AGILE
payload, using a collimated $\mathrm{^{22}Na}$ radioactive source.
By these measurements the bars physical parameters such as the
light output and the light attenuation coefficients were obtained.
After the instrument integration, MCAL was tested at the
DA$\mathrm{\Phi}$NE accelerator Beam Test Facility in Frascati
during the GRID calibration session. MCAL was then calibrated
after satellite integration at the CGS facility in Tortona,
exposing the instrument to an uncollimated $\mathrm{^{22}Na}$
radioactive source placed at different positions with respect to
the satellite axis in order to evaluate the MCAL efficiency and
the overall contribution of the spacecraft volumes to the detector
response.

The MCAL response to impulsive events and GRBs was studied by
reproducing the conditions for GRB events of durations between
30~ms and 2~s by means of a dedicated setup mainly based on moving
a radioactive source behind  a collimator; the speed of the source
determining the duration and rise time of the Burst
\citep{Fuschino2008}. With this setup all time windows of the
burst search logic above 16~ms have been stimulated and tested.

\subsubsection{The Anticoincidence system calibration}

An excellent anticoincidence system is required for an efficient
background rejection of the AGILE instrument. The AC flight units
(scintillator panels and photo-multiplier assemblies) have been
extensively calibrated at a dedicated run at the CERN PS T9 beam
line facility during the month of August 2004 (see \cite{perotti}
for an detailed presentation). The particle detection inefficiency
of the top-panel and lateral-panels plus their analog FEE was
measured to be better than $6\times 10^{-6}$ and $4\times
10^{-5}$, respectively. Adding the effect of the FEE
discrimination system somewhat increases the inefficiencies. The
final result is that the AGILE AC inefficiencies  are measured to
be below the required value of $10^{-4}$ for all panels.



\section{The AGILE satellite}


\begin{figure}[t!]
\begin{center}
    \includegraphics [height=6.5cm]{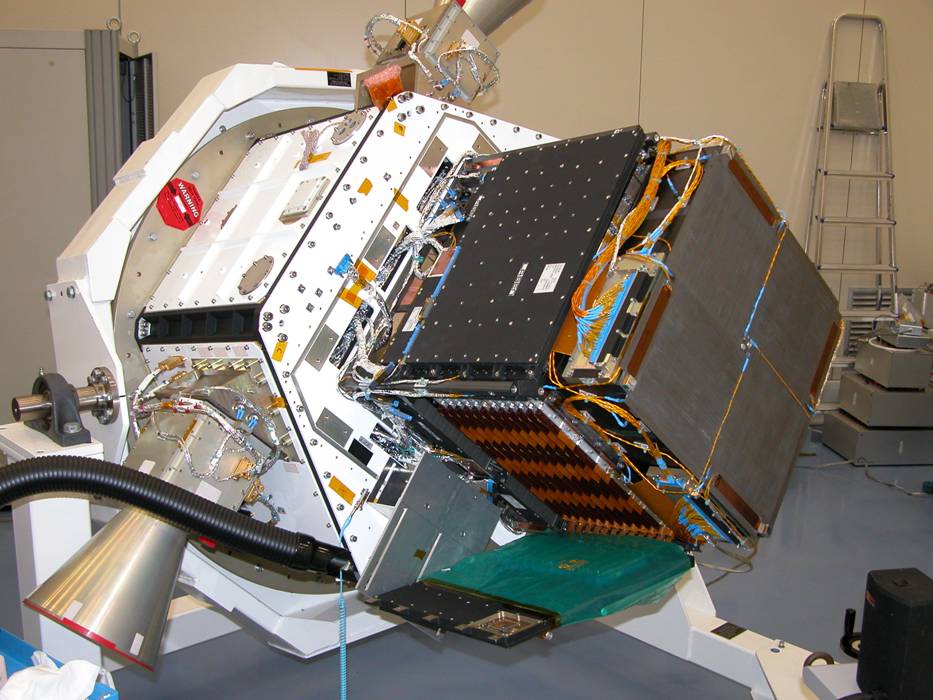} \caption { {\itshape
    The AGILE "Payload Unit" consisting of the integrated instrument
    (here shown without the anticoincidence system during integration tests)
    and the upper part of the satellite (November, 2006). }  \label{agile-sat-2}}
\end{center}
\end{figure}

\begin{figure}
\begin{center}
    \includegraphics [height=8cm]{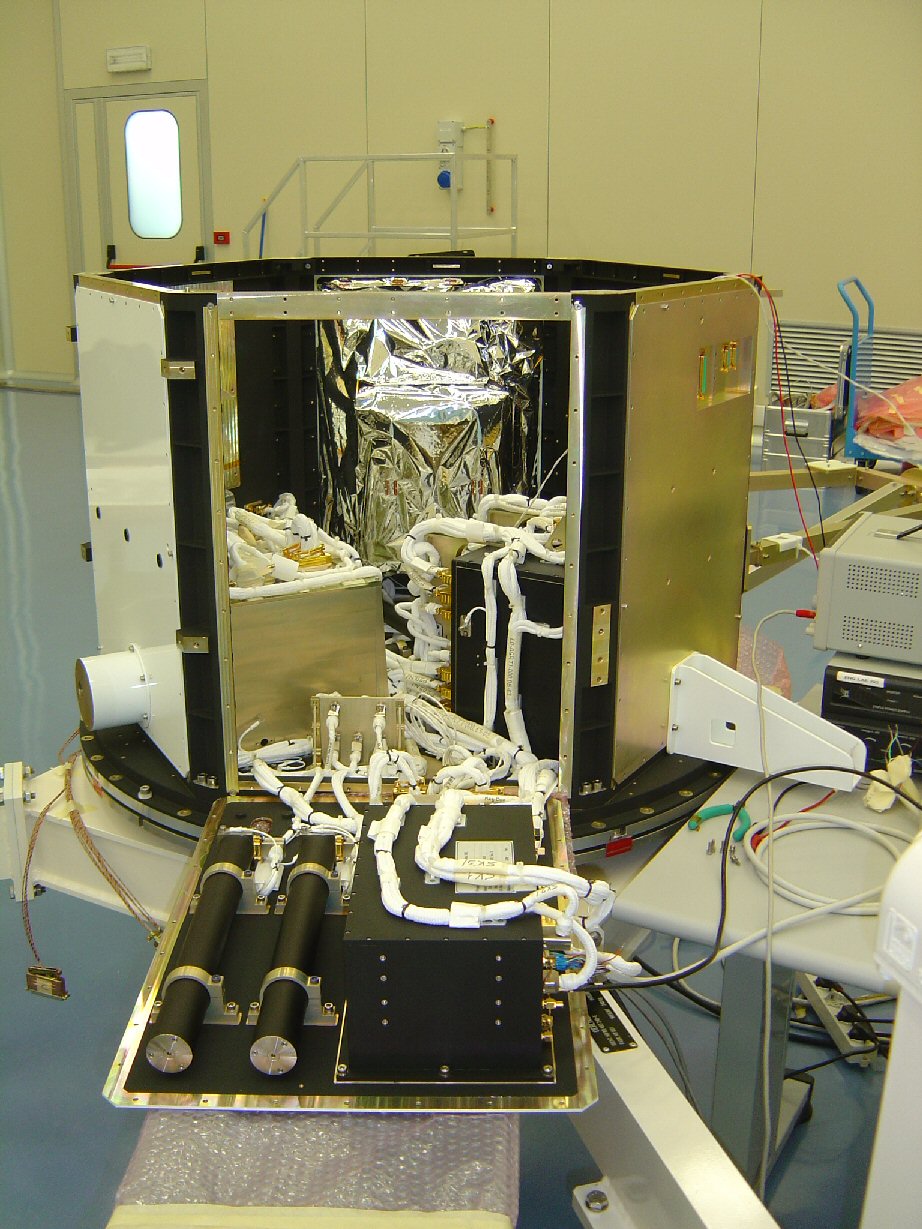} \caption { {\itshape
    The AGILE platform during the integration tests in Tortona (Italy). } }
\end{center}
 \label{bus_foto}
\end{figure}

\begin{figure}    
\begin{center}
    \includegraphics [height=8.5cm]{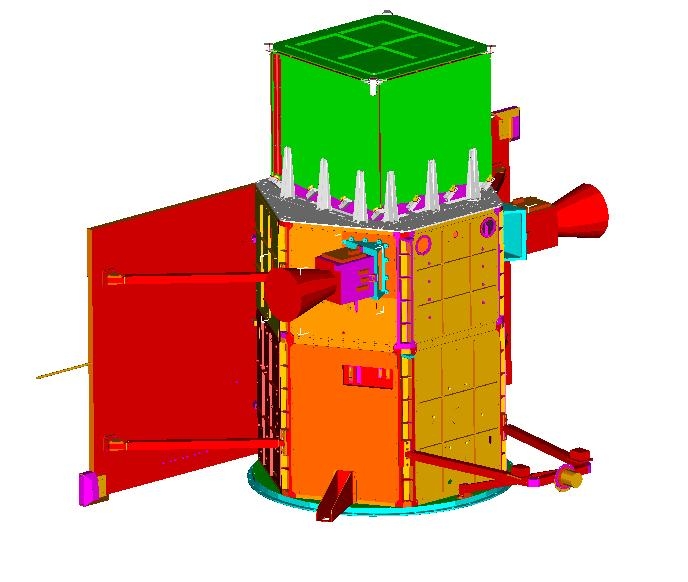} \caption { {\itshape
    A schematic view of the AGILE satellite configuration. } }
\end{center}
 \label{agile_general}
\end{figure}

\begin{figure}[h!]
\begin{center}
    \includegraphics [height=6.5cm]{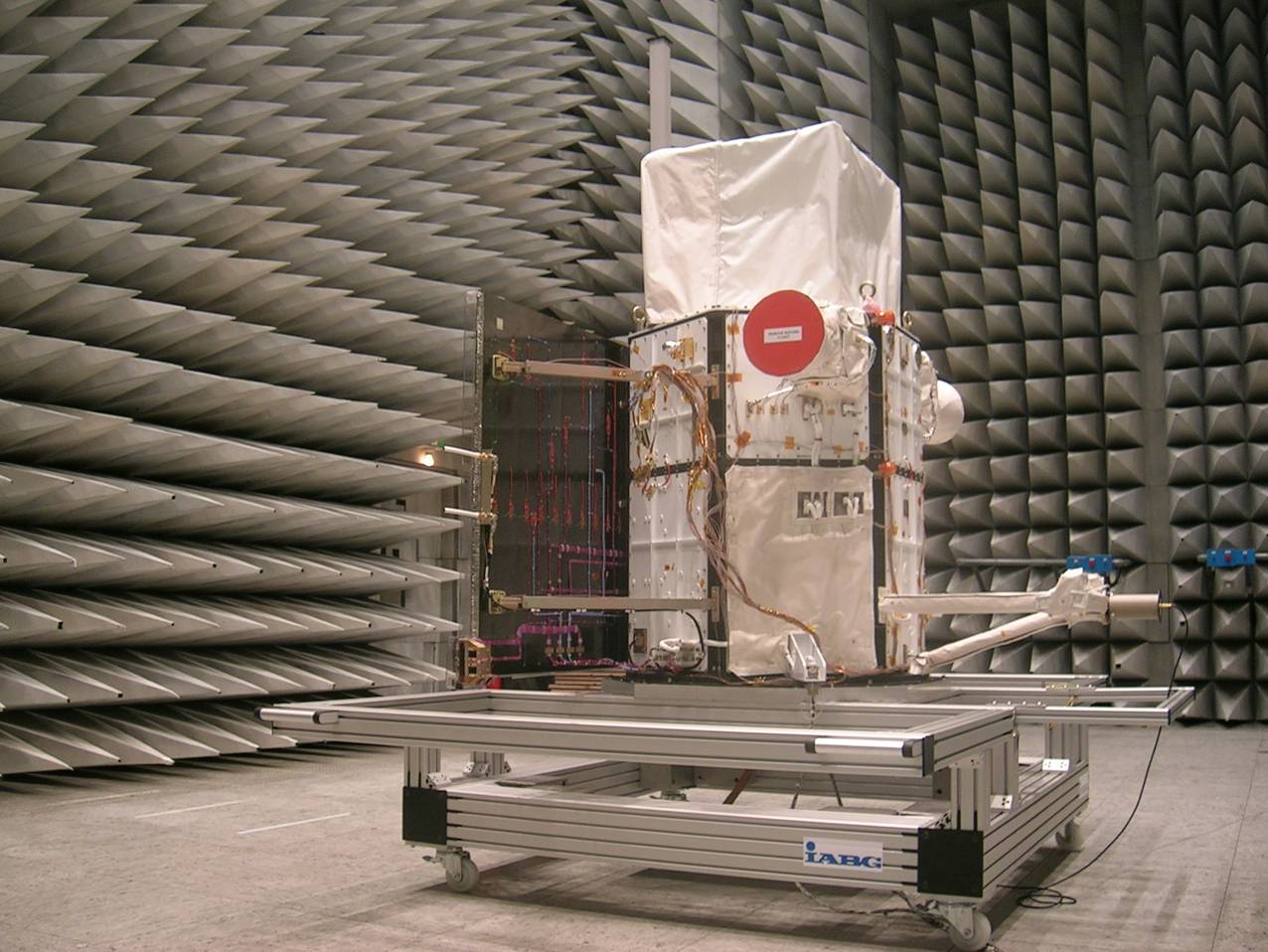} \caption { {\itshape
    The AGILE satellite during the qualification tests in IABG (Munich) (June, 2006). } }
\end{center}
 \label{agile-qualif-03}
\end{figure}

The AGILE spacecraft is of the MITA class and has been developed
by Carlo Gavazzi Space (CGS) as prime contractor
 and by Rhein-Metall (formerly Oerlikon-Contraves).
The spacecraft has been customized for the AGILE program, and
represents a very good compromise between the mission technical
requirements and the stringent volume and weight constraints.


The overall configuration has an hexagonal shape (for volume
optimization and heat dissipation requirements) and is divided
into two units:  (1) the "Payload  Unit"  (see
Fig.~\ref{agile-sat-2}) hosting the Instrument, its electronics,
the instrument DH computer unit, the Star Sensors and electronics,
the navigation unit and electronics; (2) a "service module"
hosting all the other vital control units and the satellite OBDH
system. The Instrument is mechanically and electrically integrated
with the upper unit.

The AGILE spacecraft provides all the required services for the
satellite operations and the AGILE instrument optimal functioning.
In particular, it provides the  power supply, communication with
the ground station, satellite attitude control and thermal control
capability. All these functionalities are managed by the On Board
Data Handling (OBDH) subsystem that also takes care of the
on-board monitoring and control activities.

To guarantee the required reliability for the whole mission
lifetime, all main spacecraft units have a full redundancy. The
redundancy management is mainly performed autonomously by the OBDH
software in order to provide the required autonomy to recover from
possible subsystem failures that can occur during the mission.

The Satellite thermal control is of passive type and is
implemented by placing the most dissipating devices close to the
externally radiating surfaces. A set of heaters, controlled by
thermostats or by software are used on critical equipments (in
particular the Li-Ion battery and sensitive parts of the
Instrument such as the AC system and the MCAL) to keep the
temperatures within  the required ranges.

\subsection{Solar panels and power system}

The satellite electrical power is provided by a fixed solar panel
of about $2 \, \rm m^2$ area, equipped with high-efficiency triple
junction GaAs cells. The solar panel system is positioned along
the satellite body in order to minimize the  secondary particle
background for the instrument. To supply power during the eclipse
periods and during the attitude acquisition phase, the power
system includes a re-chargeable Lithium-Ion battery with a
capacity of 33~Ah. The battery charging activity, the power
conversion and distribution to the satellite subsystems is
provided by a power unit controlled by the OBDH through a
dedicated software.

The fixed solar panels configuration constrains the AGILE pointing
strategy. The Sun direction is required to be always
quasi-orthogonal (within 1-3 degrees) to the solar panel surface.
This implies that AGILE cannot point towards the Sun:
 most of the sky is accessible by the large FOV AGILE
gamma-ray imager but the solar and anti-solar directions are
excluded from direct pointings.

\subsection{Satellite attitude control}

The AGILE satellite attitude control system (ACS) is designed to
achieve the  scientific requirement of the satellite attitude
reconstruction of $\sim$1 arcmin. The AGILE ACS uses a set of
sensors and actuators controlled by a dedicated software running
on the OBDH computer to guarantee the required attitude pointing
in all the mission phases. After the launcher separation, the ACS
was able to acquire, within a few orbits, a Sun pointing attitude
to guarantee the necessary power generation from the solar array.
Once acquired, the Sun pointing attitude is maintained for the
whole mission using two different control algorithms: (1) a
"coarse" Sun pointing attitude, that makes use of a reduced set of
sensors and actuators to maintain the orthogonality of solar
incidence on the solar panel for a free-wobbling of the satellite
pointing; and (2) a "fine" Sun pointing attitude, using the whole
on-board capability to ensure an overall satellite pointing
accuracy better of 1 degree and the required attitude stability
better of 0.1 degrees/s. The ACS ensures the pointing of  the
payload instrument towards a given direction always  maintaining
the Sun-satellite direction orthogonal to the solar panel surface.
For AGILE, the near-orthogonality of the solar panel surface and
the Sun-satellite direction is enforced within a few degrees of
tolerance.

The AGILE satellite meets the 1-2 arcmin attitude reconstruction
requirement by using  two back-to-back oriented Star Sensors
(S/Ss) which are then an essential part of the satellite ACS. The
star sensors (manufactured by Galileo Avionica) are equipped with
on-board sky charts and software able to determine the satellite
attitude. This information is used to obtain on board the sky
coordinates for the gamma-ray individual events and for the
Super-AGILE imaging system. To minimize the grammage surrounding
the gamma-ray instrument, the S/Ss are positioned  in a
back-to-back configuration in the "PL Unit" behind the solar panel
and relatively far from the payload.

\subsection{Satellite timing and orbital positioning}

The AGILE satellite is equipped with a  navigation transceiver
that ensures an on-board timing accuracy within 2 microseconds.
The navigation system information is first used on-board by the
instrument DH unit to time tag individual events of the various
detectors. This  telemetry is also used by the Ground Segment to
improve the satellite orbital tracking.

\subsection{Satellite communication and telemetry}

Communication with the ground station is ensured by an S-band
transceiver that contacts the ASI ground station located in
Malindi (Kenya) about 14 times a day, for a total visibility
periods of about 160 minutes per day. During the visibility
periods the S-band transmitter downloads the payload data,
previously stored in the OBDH mass memory, with a net data rate of
500 kbps. The S-band transceiver is also used to upload  the
configuration telecommands necessary to the satellite operations.

The AGILE scientific instrument generates under normal conditions
a telemetry rate of $\sim 50$~kbit/s. The satellite downlink rate
for scientific telemetry is $512 \; \rm kbit \, s^{-1}$. This rate
 is adequate to transmit at every passage over the ground station
all the satellite and scientific data.

The AGILE satellite provides also a special communication link
dedicated to the fast transmission to the scientific community of
basic information related to the on-board detection and processing
of Gamma-Ray Bursts. This functionality is obtained using an
ORBCOMM transceiver controlled by the spacecraft OBDH. The ORBCOMM
transceiver is constantly connected to the ORBCOMM constellation
(composed by 36 satellites in low Earth orbits) that is able to
send a dedicated e-mail message to selected users as soon as the
Gamma-Ray Burst is detected, independently from the baseline
S-band ground station coverage. The AGILE ORBCOMM message to the
ground  provides the timing, coordinate reconstruction, and peak
flux of the transient event.

\section{The Satellite Qualification Campaign}

The AGILE satellite carried out an extensive qualification
campaign during the period June-July, 2006, with additional
testing period during February-March, 2007. All AGILE satellite
qualification tests were carried out at the IABG facility in
Munich (Germany).

During the first campaign, an extended set of measurements and
tests were carried out, including  satellite mass and inertial
measurements, acoustic tests, mechanical vibration tests, EMC
tests, thermal vacuum,  and thermal balance tests.

All the qualification tests were defined and performed according
to an environmental test plan based on the launcher and orbit
characteristics. In particular, thermal vacuum cycles were
performed in the temperature ranges (-20, + 40) for the operative
state, and (-30, +50) for the non-operative state.

After all major individual tests, specific reduced functional
tests were performed for both spacecraft and instrument. At the
end of the qualification campaign, the test results were presented
and discussed at the AGILE Test Review Board meeting.

Independently of the qualification outcome, for programmatic
reasons the AGILE satellite was subject to an additional reworking
during the period September-November, 2006. A significant
reassembling of parts of the satellite was necessary to implement
the substitution of a few components. An additional satellite
qualification test phase was then carried out in February, 2007
until the end of March, 2007. A complete satellite functional test
was performed in order to verify the nominal operation of
spacecraft and instrument subsystems including OBDH redundancy
testing.

\begin{figure}[t!]
\begin{center}
    \includegraphics [height=6.5cm]{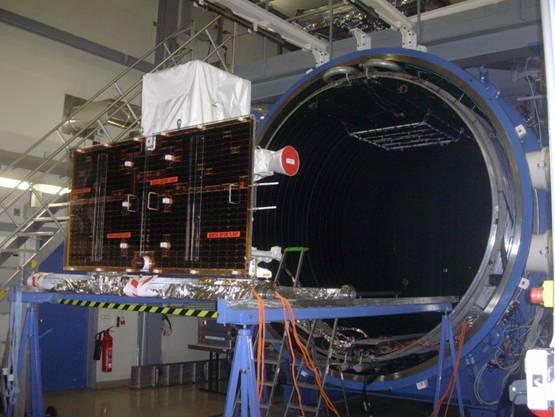} \caption { {\itshape
    The AGILE satellite being moved into the thermal-vacuum test chamber in IABG (Munich) (July, 2006). } }
\end{center}
 \label{agile-qualif-02}
\end{figure}

 \begin{figure*}
 \begin{center}
 \includegraphics [height=8.5cm]{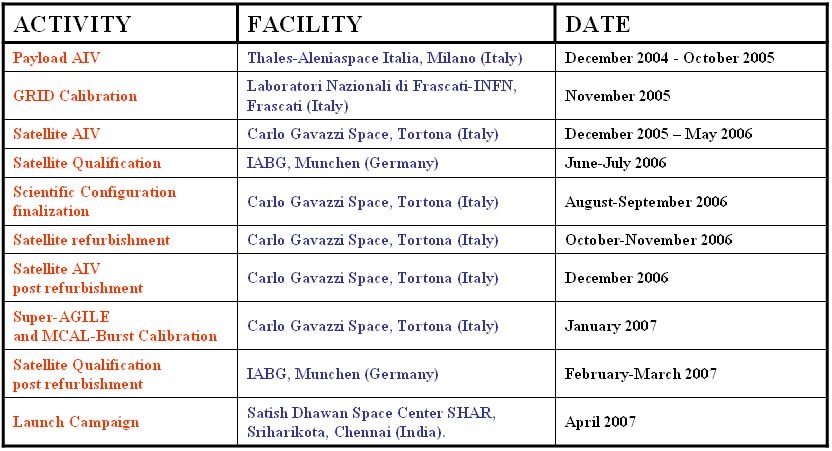} \caption { {\itshape
     Summary of the AGILE instrument functional and scientific performance test runs
    carried out during the AIV, calibration and qualification phases. } }
 \end{center}
  \label{agile-qualif-01}
 \end{figure*}

\section{The pre-launch AGILE scientific performance}

The scientific performance of the AGILE instrument has been
checked during all the payload and satellite integration and
testing phases. The tests carried out at the end of specific
qualification sessions confirmed the nominal behavior of the
instrument subsystems.

After having completed the satellite qualification tests, a series
of additional final scientific measurements were performed, and
their results were presented  at the Mission Review Board. All
detectors and subsystem functioning was judged to be nominal and
within the scientific requirements and specifications.

\begin{figure}
\begin{center}
    \includegraphics [height=6.5cm]{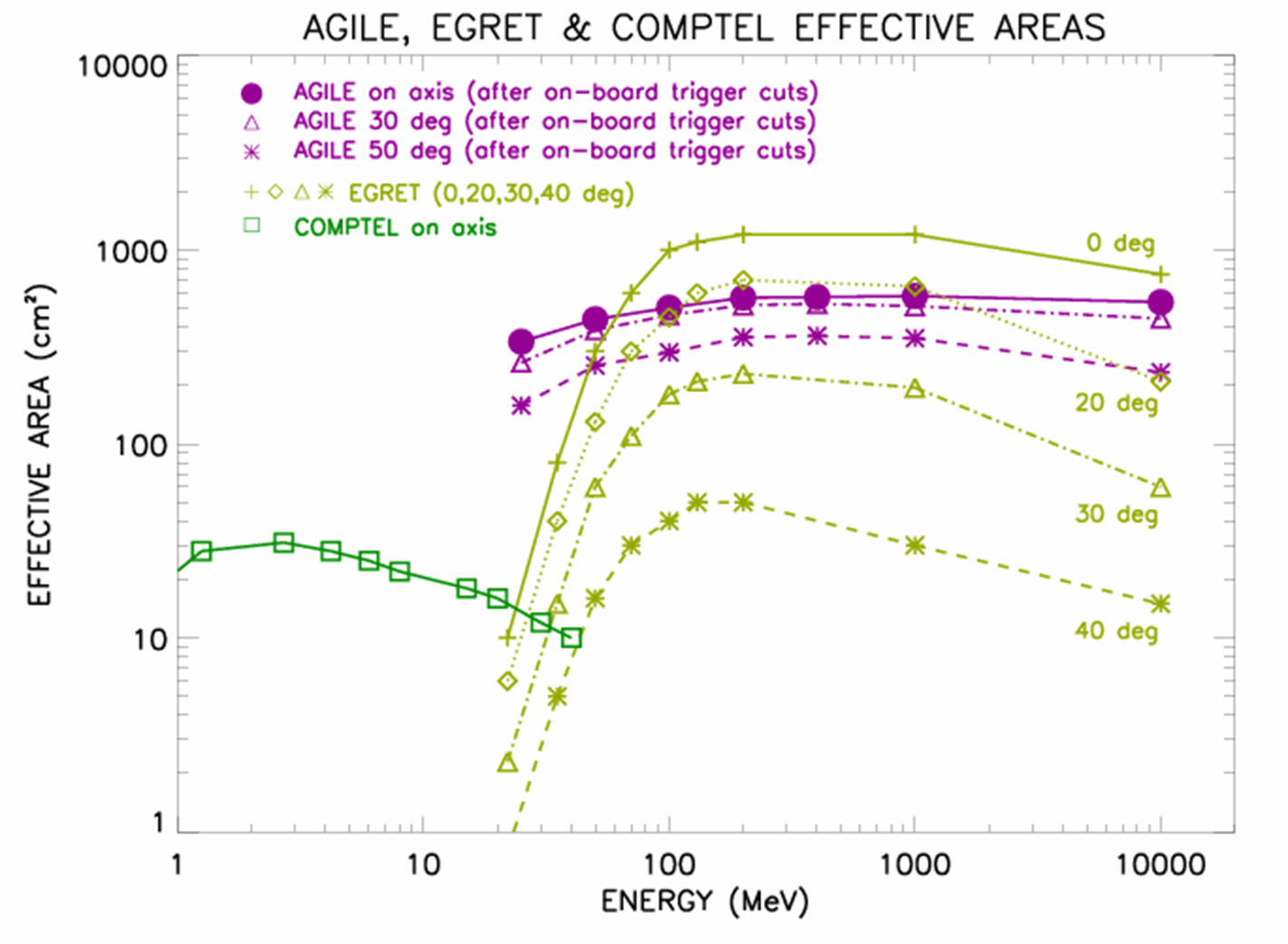} \caption { {\itshape
    The AGILE Gamma-Ray Imaging Detector (GRID) effective area as obtained by
    pre-launch simulations and instrument tests. } \label{agile-science01}}
\end{center}
\end{figure}

\begin{figure}[t!]
\begin{center}
\includegraphics [height=6.5cm]{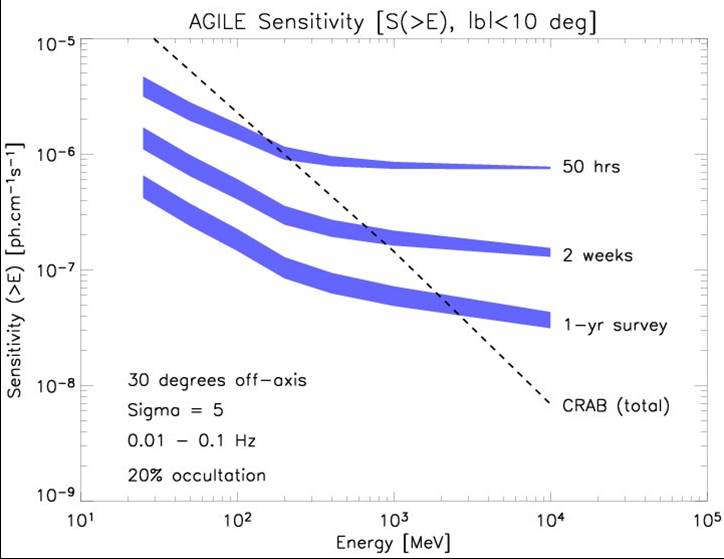} \caption { {\itshape
    Simulated integrated flux sensitivity of the AGILE-GRID
     as  a function of energy for a 30-degree off-axis source in the
Galactic plane. The Crab spectrum is shown by the dotted line. }
\label{agile-science02}}
\end{center}
\end{figure}

\begin{figure}[h!]
\begin{center}
\includegraphics [height=8.5cm]{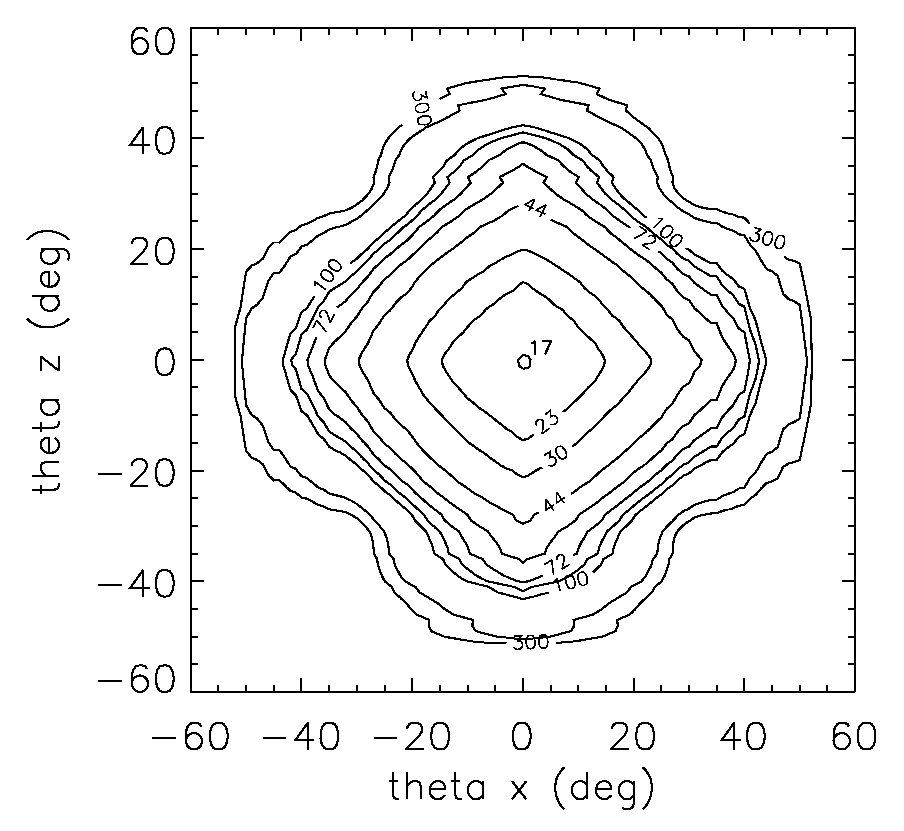} \caption { {\itshape
    Simulated sensitivity of Super-AGILE detector (all four units)
    expressed as (5 sigma) contour levels (in mCrab) over the Field of View in the energy range
    20-55 keV.
    Calculation carried out for a Crab-like spectrum, 50 ksec
    exposure.   } \label{SA-sens}}
\end{center}
\end{figure}

\begin{figure}[h!]
\begin{center}
 \includegraphics [height=6.5cm]{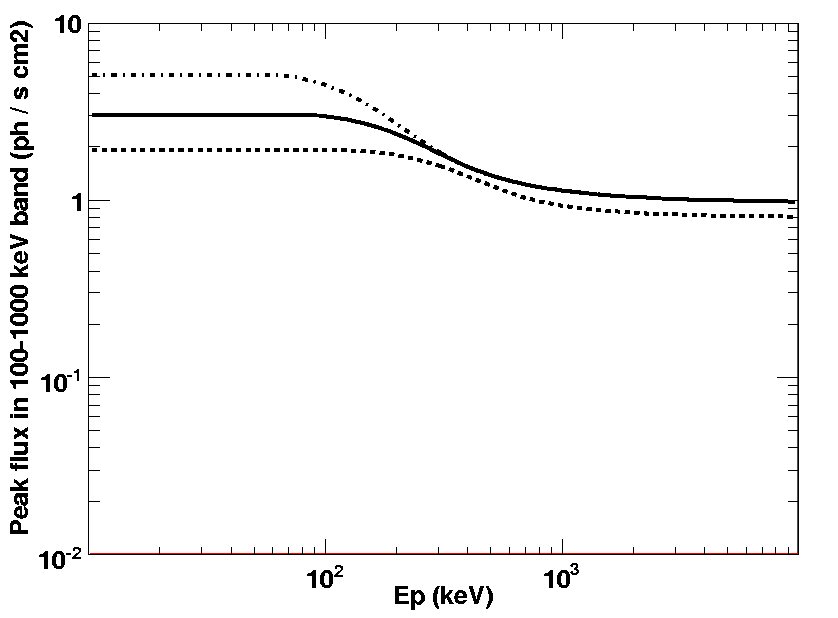} \caption { {\itshape
 MCAL sensitivity to GRBs vs. peak energy at 30 deg
off-axis, for three different GRB spectra modeled according to the
Band model \citep{Band1993}. Continuous line: $\alpha=-1$,
$\beta=-2.5$; dashed line $\alpha=-0.5$, $\beta=-2$; dot-dashed
line $\alpha=-1$, $\beta=-3$.} \label{mcal-sens}}
\end{center}
\end{figure}

The pre-launch overall scientific performance of the AGILE
instrument detectors in terms of confirmed effective areas and
sensitivities is summarized in Figs. \ref{agile-science01},
\ref{agile-science02}, \ref{SA-sens}, and \ref{mcal-sens} for the
GRID, Super-AGILE and MCAL detectors, respectively.

\section{The AGILE Ground Segment: Mission Operation Center}

The AGILE ground segment is an essential part of the Mission. It
is divided in three components under ASI supervision and
management:

\begin{itemize}

\item   the ASI communication ground base in Malindi (Kenya);

\item   the Mission Operations Center (MOC), located at the
Telespazio facility in Fucino (Italy);

\item  the AGILE Data Center, located at the ASI Science Data
Center (ASDC) in Frascati (Italy).

\end{itemize}

In this Section, we describe the AGILE Ground Segment (AGS)
devoted to in-orbit operations. The AGILE Scientific Data Center
is described in Section~\ref{ADC}.

\begin{figure*}   
\begin{center}
  \includegraphics [height=10cm]{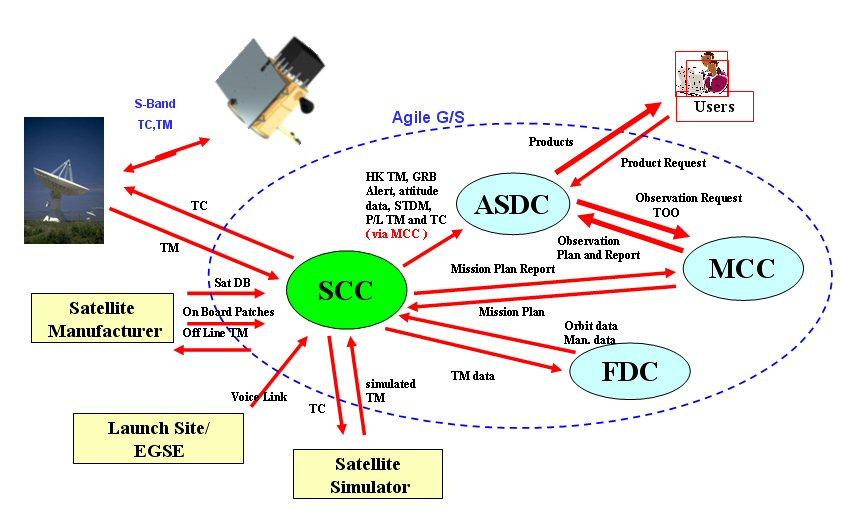} \caption { {\itshape
    A schematic view of the AGILE Ground Segment. } \label{GS}}
\end{center}
\end{figure*}

The  satellite operations are carried out by the Ground Segment
unit of Telespazio under the supervision of the ASI Mission
Director. The AGILE Ground Segment  is the main system in charge
of satellite monitoring and control. AGS is also in charge of the
 instrument observation planning and execution upon specific
requests by the AGILE Team. The AGS main functions can be
summarized as follows:
\begin{itemize}
\item satellite tracking
and acquisition during the  14 passes per day over the TT\&C
Malindi ground station;
\item TC \& TM handling (TM acquisition,
processing, display, archiving; TC generation, verification and
uplink);
\item
 satellite orbit determination and propagation, satellite
attitude determination, support to satellite ACS on-board
operations; \item satellite sub-systems and payload monitoring and
control, satellite modes of operations monitoring and control;
\item
mission planning generation and execution.
\end{itemize}

The instrument  is continuously monitored and controlled with the
bus satellite sub-systems.   Scientific data are received by the
Malindi ground station, transmitted to the Mission Operations
Center (MOC) ground station in Fucino, and locally archived there.
Subsequently, typically within a few minutes, scientific data are
transmitted to the ASI Scientific Data Center in Frascati. The
AGILE Mission Operations Center is composed of the following main
sub-systems: (1)
 the Satellite Control Center (SCC); (2) the Flight Dynamics
Center (FDC); (3) the Mission Control Center (MCC); (4) the TT\&C
ground station and communication network. Fig.~\ref{GS} shows a
schematic summary of the AGILE GS.

The AGILE Satellite Control Center (SCC), located in Fucino, is in
charge of all the satellite monitor and control functions, both in
nominal and in contingency situations. The main SCC functions are
the following: (1) satellite telemetry acquisition from TT\&C
ground station, TM processing, display and archiving; (2) handling
of the satellite acquisition automatic procedures; (3) satellite
database handling and maintenance; (4) satellite sub-systems and
payload health monitoring and control through housekeeping data
presented on alpha-numeric and mimic displays; (5)   real time or
time-tagged telecommand preparation, TC to be uplinked to the
satellite; (6) co-ordination with the TT\&C ground station for
data and voice exchange and with reference to program track mode
antenna operations and to satellite tracking data reception to be
processed by FDC; (7) communication network management.

The AGILE Flight Dynamics Center (FDC) carries out the satellite
orbit determination and prediction,  obtains the satellite
attitude dynamics determination, and supports the satellite AGS
on-board operations. The  main FDC tasks are: (a) satellite
performance monitoring; (b) attitude maneuver calculation for the
pre-defined satellite pointing at scientific targets; (c) star
tracker and navigation system receiver data handling. The AGILE
FDC  generates standard products including: (1) the Orbital Files,
necessary for the satellite in-orbit operations and for the
mission planning activity; (2) the Attitude Files, containing  the
reconstructed satellite attitude, necessary for the satellite
in-orbit operations and for scientific activities support; (3)
files containing the needed information to prepare the payload
configuration tables to be uploaded (contact tables, SAA tables,
occultation table, Earth vector table, Star Trackers table); (4)
files containing the information on parameters necessary to
support the spacecraft activity; (5) the Orbital File for
scientific purposes; (6) a Sequence of the Events file; (7)  Two
Line Elements (TLE) to be used for the TT\&C ground antenna
Program Track mode.

The AGILE orbit determination is based on navigation system
telemetry data processing and, as a backup, on AMD (Angular
Measurement Data) measurements stored at TT\&C ground station with
the antenna in auto-track mode.

The AGILE Mission Control Center (MCC), performs the mission
planning and the instrument data handling management. In
particular, the MCC is in charge of: (1) the Payload scientific
raw data archiving (level 0 archiving) and delivery to ASDC; (2)
the Mission planning generation and preliminary check-out,
according to the scientific observation requests; (3) providing
full support for both the payload planning and the creation of
command procedures to perform the AGILE flight operations.

The AGILE TT\&C ground station, based at the ASI Malindi (Kenia)
Broglio Space Center, is in charge of S-band (S TX 2025 -2120 MHz
e RX 2200-2300 MHz) RF ground to space satellite communication
during all the AGILE mission phases. The main TT\&C ground station
tasks are the following: (1)   ground to space RF S-band interface
with the satellite during all the mission phases; (2) during
satellite visibility over the station, telemetry is received from
the satellite and telecommands (TCs) are uplinked to the
satellite; (3) real-time TM is extracted and sent to SCC, while
off-line TM is locally stored to be sent to SCC at the end of the
satellite pass; (4) TCs received from SCC are uplinked to the
satellite for immediate or time-tagged on-board execution; (5)
during satellite visibility over the station, the satellite
tracking is performed, in 'program track' or 'auto-track' modes of
operations; (6) ground station equipment and configuration monitor
and control.

The AGILE Communication Network, mainly based on the ASINET
network provided by ASI, carries out the data and voice
communication between the AGILE GS sub-systems during all the
mission phases. The AGILE Data Center located at ASDC is the ASI
official interface with the AGILE GS during the satellite
operations. It interacts with the MOC  in submitting scientific
requests of observations, and  receives from the  AGS the AGILE
payload raw data. In case of a GRB event it also receives
notification.

The equatorial orbit of the AGILE satellite leads to a satellite
visibility every 90 minutes for a period of about 10 minutes.
During each pass, the satellite receives telecommands from the
Ground Station and downloads housekeeping and scientific
telemetry. The MOC carries out the operational management of the
satellite and the ground segment, guaranteeing the pre-defined
observation plan provided by the AMB, and also guaranteeing the
delivery of the Scientific data to ASDC.

In order to guarantee the correct management of the in-orbit
operations with respect to the Mission Safety requirements, the
MOC defines  Operational Procedures and Operational Strategies
allowing to command the satellite operations for  nominal and
contingency phases.  The sequence of commands, prepared by the
Mission Control Center (MCC), are to be sent in a  "time-tagged"
format, using the satellite on-board storage ability. In case of
on-board anomalies, the satellite is able to self-configure in a
"safe" state that guarantees its surviving for at least 72 hours.

\begin{figure*}  
\begin{center}
    \includegraphics [height=12cm]{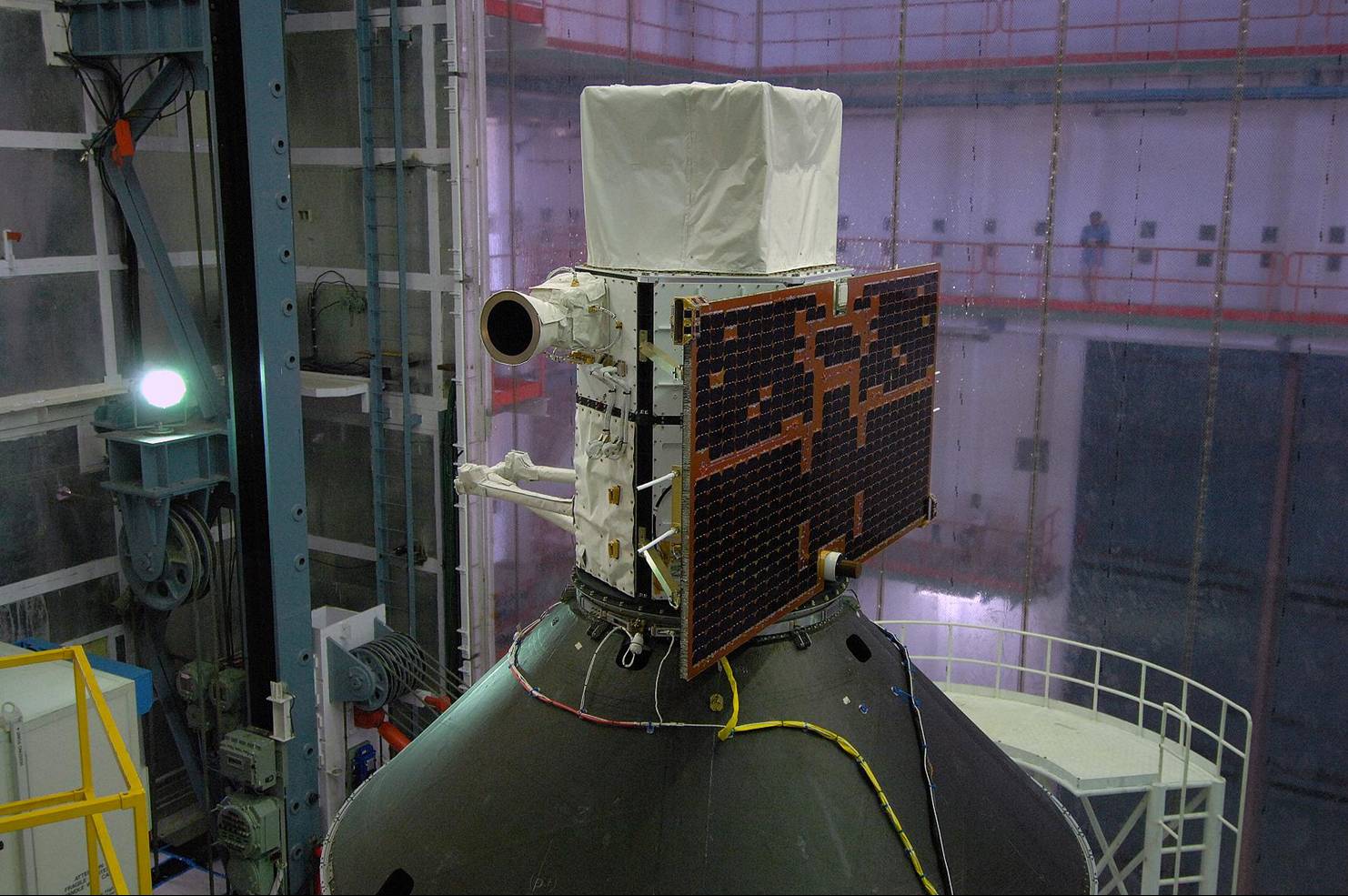} \caption { {\itshape
    The AGILE satellite integrated with the fourth stage of the PSLV C-8
    rocket in the ISRO Sriharikota launch base (April 15, 2007). } }
\end{center}
 \label{sat-2}
\end{figure*}

\section{The Launch Campaign and Orbital Parameters}

The satellite was transferred  from Munich (Germany) to the ISRO
Sriharikota base near Chennai (India) at the end of March, 2007.
The AGILE launch campaign in Sriharikota started on April 1, and
ended with a very successful  launch on April 23, 2007.

\subsection{The launch campaign}

A series of functional and reduced scientific tests were carried
out in the Sriharikota satellite integration facility during the
period April 2-9, 2007. A rehearsal of the complete test procedure
of the in-orbit Commissioning of the satellite was  performed.
These tests included the execution of all baseline satellite
in-orbit procedures. Spacecraft and instrument commandability were
extensively checked.

The AGILE satellite integration with the PSLV-C8 launcher started
on April 12 and was successfully completed on April 17, 2007. At
the end of this integration phase and before the final fairing
assembly completion, a reduced set of functional and scientific
tests were performed. The integrated PSLV-C8 rocket was
transported to the launch pad
 on April 18, 2007: the launch countdown started on April 20, 2007.

\begin{figure}  
\begin{center}
    \includegraphics [height=6cm]{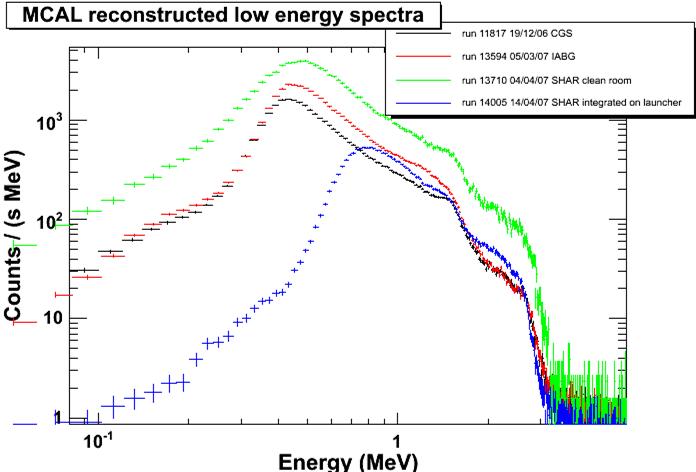} \caption { {\itshape
An example of the AGILE pre-launch scientific measurements carried
out before and during the AGILE launch campaign. The plot shows a
sequence of spectral measurements obtained with the AGILE MCAL for
different environmental and testing conditions. {\sl Black data
points}: data obtained during the satellite integration at the CGS
facility in Tortona (Italy). {\sl Red data points}: data obtained
during the satellite qualification at the IABG facility in Munich
(Germany). {\sl Green data points}: data obtained during the
satellite launcher pre-integration tests at the Sriharikota
facility. {\sl Blue data points}: data obtained during the final
pre-launch tests with the satellite completely integrated with the
PSLV-C8 at a height of about 50 meters above ground. In this case,
the MCAL threshold was set near 600 keV. Note the difference of
the pre- and post-integration spectra interpreted as caused by
different natural radioactive background conditions.
 } }
\end{center}
 \label{mcal-2}
\end{figure}

 On April 23, 2007, at 12:00 noon local time, the PSLV-C8 rocket
 was successfully launched with a nominal performance. After about
 20 minutes, the fourth stage (with the AGILE satellite still
 integrated) performed the inclination correction maneuver
 and achieved the nominal orbit for the mission with the
 required height and inclination. Satellite and
fourth stage separation occurred immediataly afterwards.

 The first contact with the Malindi ground station occurred
 nominally at the first satellite first pass over Kenya, about 75 minutes
 after launch.

\subsection{Orbital parameters}

The AGILE orbit is quasi-equatorial, with an inclination of 2.5
degrees and average altitude of 535 km.
 A small inclination low earth orbit (LEO) is a clear plus
 for our mission because of the reduced particle
background (as immediately verified in orbit by all instrument
detectors). Furthermore, a low-inclination orbit optmizes the use
of the ASI communication ground base at Malindi (Kenya).
Table~2  
provides the final AGILE orbital parameters. Depending on solar
activity, the AGILE satellite re-entry in the atmosphere is
predicted to happen not earlier than 2012.

\begin{table}
\large
\begin{center}

{\bf Table 2: AGILE Orbital Parameters}

\vspace*{.2in}
\begin{tabular}{|l|l|}
  \hline
  Orbital parameter &  \\
  \hline
  Semi-major axis &  6,922.5 km \\
  Average altitude & 535 km \\
  Eccentricity &  0.002 \\
  Inclination angle &  2.47 degrees\\
  Apogee altitude & 553 km \\
  Perigee altitude & 524 km \\
\hline
\end{tabular}
\end{center}
 \label{table-2}
\end{table}

\section{Early operations in orbit}

After the nominal launch and the correct satellite attitude
stabilization within approximately 2 days, the operations focused
on two different tasks: (1) the commissioning of both the
satellite platform and  instrument; (2) in-orbit scientific
calibration of the instrument. We briefly describe here these two
main activities, postponing a detailed description to forthcoming
publications.

\subsection{The satellite in-orbit Commissioning and instrument
checkout phase}

The satellite platform was tested and functionally verified in all
its main capabilities during the last days of April. The checkout
sequence of tests ended with the satellite fine pointing attitude
finalization that implies the nominal $\sim$1 degree pointing
accuracy and the 0.1 degree/sec stabilization. Attitude
reconstruction, both on-board and on the ground, was tested to be
initially within a few arcminute accuracy. A sequence of early
pointings was carried out, typically lasting for a few days.

\begin{figure}[t!]
\begin{center}
    \includegraphics [height=7.0cm]{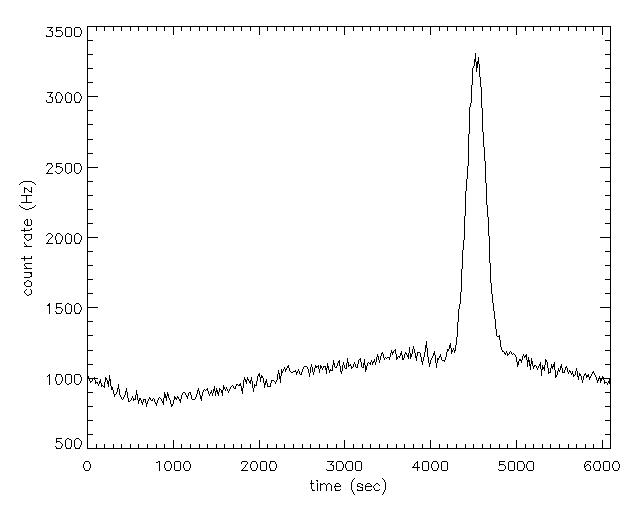} \caption { {\itshape
AGILE AC count rate throughout the whole orbit for one of the 12
lateral Anticoincidence panels. The count rate peak occurs in
coincidence with the passage over the South Atlantic Anomaly. }
\label{agile-AC}}
\end{center}
\end{figure}

The instrument subsytem switch on started in early May and
proceeded with nominal behavior of all detectors. A first check of
the instrument housekeeping telemetry indicated a nominal particle
background rate as predicted by extensive simulations
\citep{cocco,longo2}. Fig.~\ref{agile-AC} shows a typical
background count rate on a lateral AC panel throughout the
equatorial orbit.

\begin{figure}  
\begin{center}
\includegraphics [height=7.5cm]{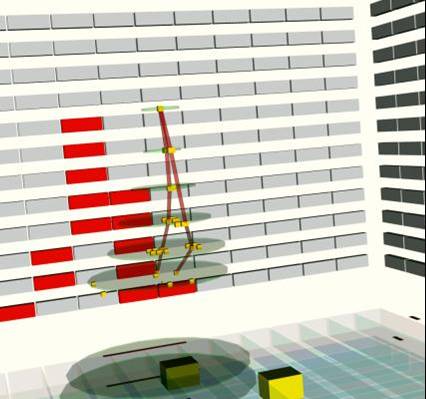} \caption { {\itshape
  The first gamma-ray event detected by AGILE in space (May
10, 2007).  } \label{fig-4c}}
\end{center}
\end{figure}

The GRID acquisition rate after a complete on-board processing and
Earth gamma-ray albedo rejection turned out to be stable
throughout the orbit with a modulation induced by the Earth
sweeping the GRID FOV (within 1-7~Hz). A detailed account of the
GRID and overall on-board scientific telemetry will be reported
elsewhere \citep{argan3}. The GRID subsystem was extensively
tested first, and the baseline trigger logic photon acquisition
started on May 10th. Fig.~\ref{fig-4c} shows the first gamma-ray
photon detected by the GRID and transmitted to the ground as the
first event of the first telemetry packet. The overall particle
background turned out  to be within the expected rate, both at the
level of the Anticoincidence rate, and especially at the so called
Level-1 and Level-2 event processing \citep{argan3}.

\begin{figure} 
\begin{center}
    \includegraphics [height=4.5cm]{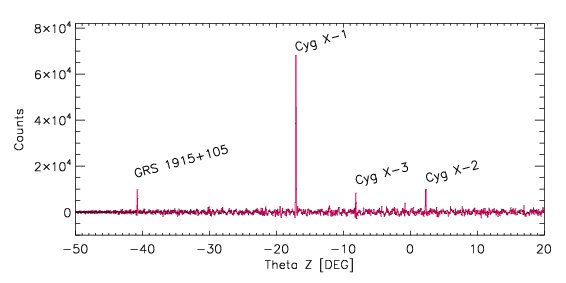} \caption { {\itshape
Super-AGILE deconvolved sky image (one detector unit) of the
Galactic plane obtained on June 2, 2008 (total effective exposure
of 45 ksec). Several hard X-ray sources are detected as marked in
the figure. } \label{agile-cyg}}
\end{center}
\end{figure}

 The Super-AGILE detector was tested
immediately afterwards with a dedicated pointing of an
extragalactic field. A detailed scan and test of the individual SA
strip thresholds was carried out with an optimized parameter
stabilization procedure \citep{pacciani}. Hard X-ray data were
obtained with a nominal performance and very low-background. These
operations ended successfully in mid-July, 2007 \citep{feroci4}.
Fig.~\ref{agile-cyg} shows an example of a typical 1-day pointing
in the Galactic plane with the detection of Cyg X-1, Cyg X-2, Cyg
X-3, and GRS 1915+105.

\begin{figure}[t!]
\begin{center}
    \includegraphics [height=6.0cm]{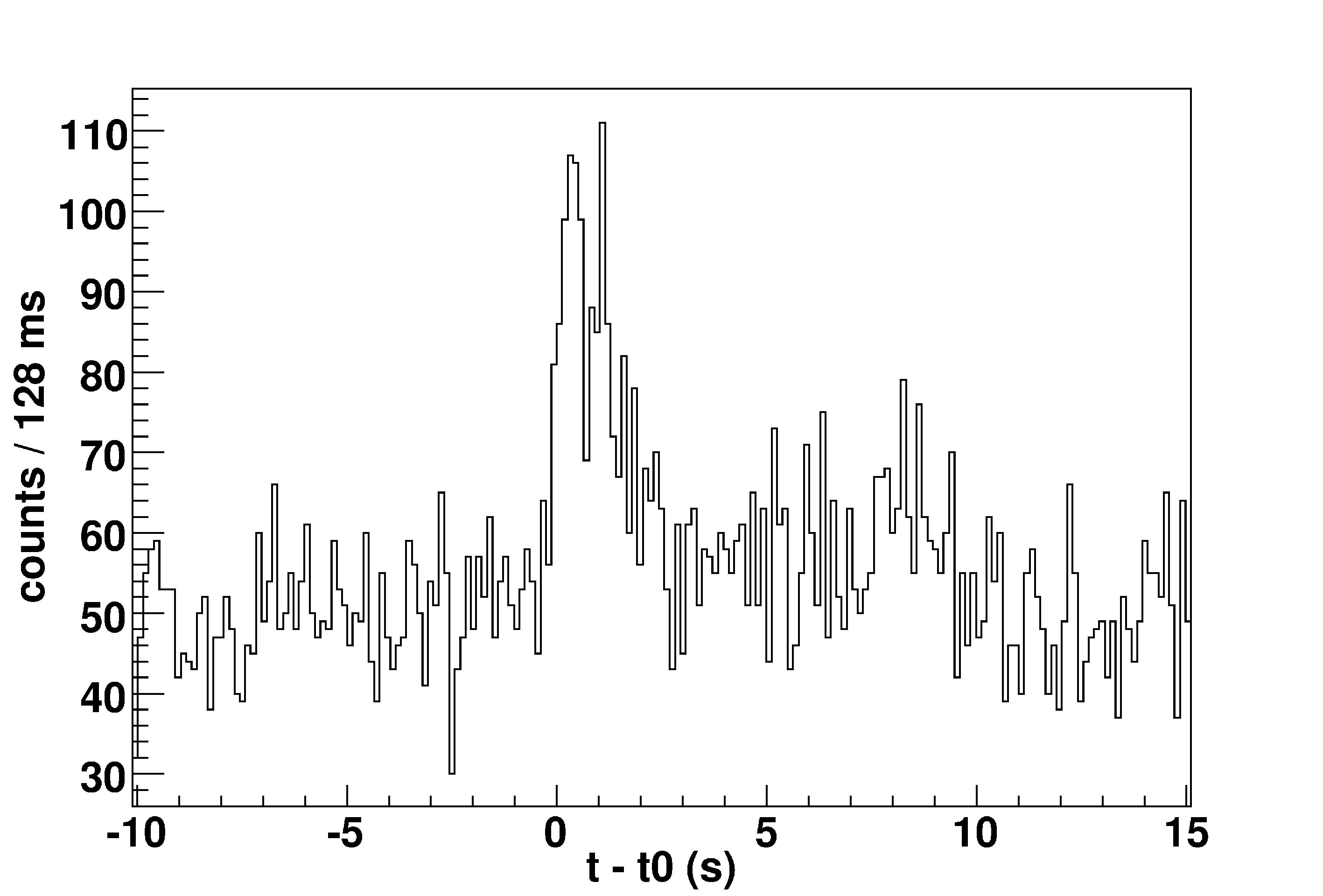} \caption { {\itshape
 The lightcurve (128 msec time bin) of GRB 080319C (Pagani et al.,
2008; Marisaldi et al., 2008) detected by the AGILE
Mini-Calorimeter in the energy band 0.3-5 MeV.} \label{agile-grb}}
\end{center}
\end{figure}

The MCAL detector optimization and configuration checkout was
performed in parallel with other instrument testing. A
satisfactory detector configuration was obtained at the end of
June, 2007 \citep{marisaldi4}. Fig.~\ref{agile-grb} shows an
example of a typical GRB detected by the MCAL in the energy range
0.3-5 MeV.

\subsection{In-orbit instrument calibration}

An account of the AGILE pointings and in-orbit calibration details
will be presented elsewhere. We summarize here the main outcomes
of the early in-orbit calibration phase.

\begin{figure}[h!]
\begin{center}
    \includegraphics [height=7.5cm]{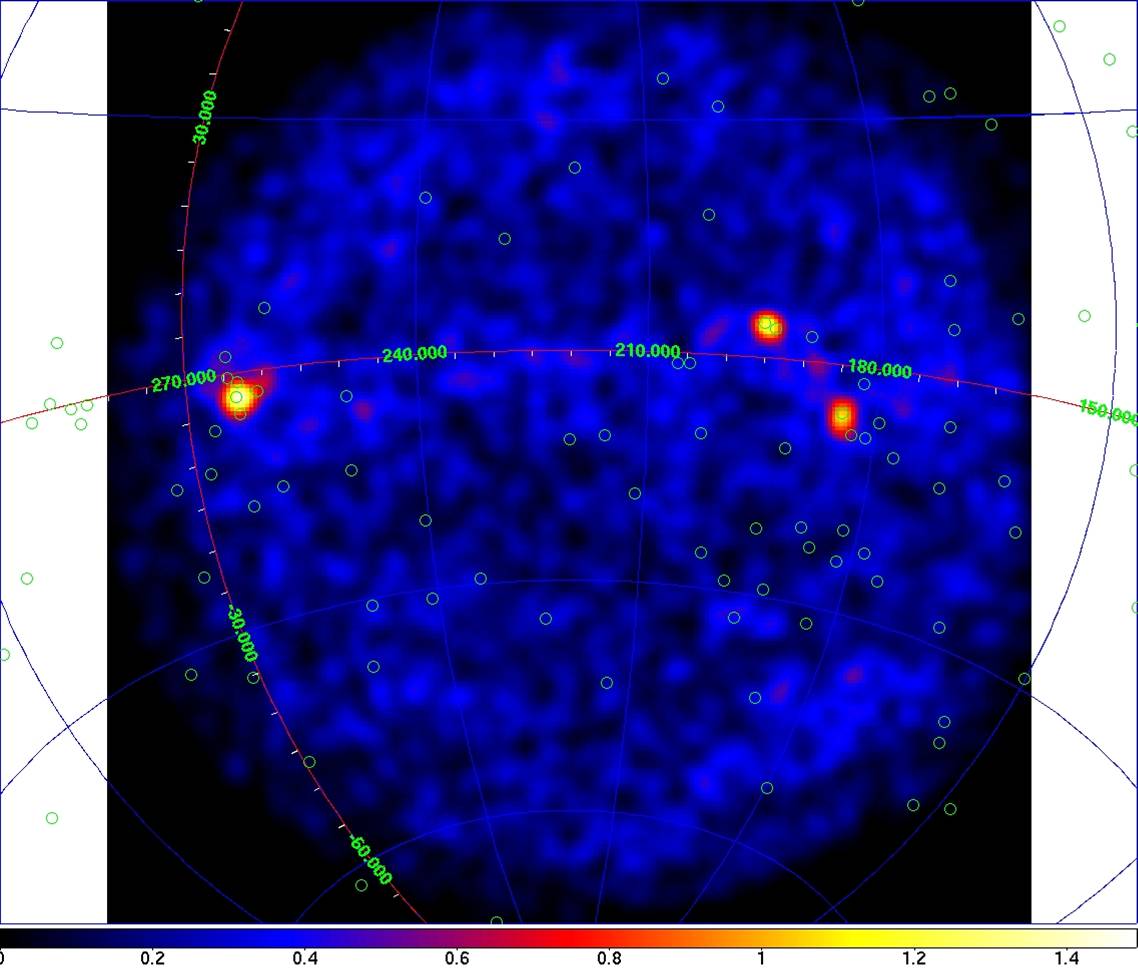} \caption { {\itshape
AGILE 1-day gamma-ray counts map for photons above 100 MeV
obtained on September 28, 2007. The large-field of view sky counts
map shows in the same picture all three most important gamma-ray
pulsars: Vela,  Crab  and Geminga. } \label{agile-science10}}
\end{center}
\end{figure}

Preliminary gamma-ray data obtained for  the Vela pulsar in early
June, 2007 confirmed immediately the good quality of the GRID
background rejection and its imaging capability. The scientific
operations started in early July, 2007 with a  2-month observation
of the Vela pulsar region. At the end of August, 2007, AGILE
devoted about 1 week to a  pointing of the Galactic Center region.
Finally, AGILE carried out the gamma-ray and hard X-ray
calibration with the Crab pulsar during the months of September
and October, 2007. Fig.~\ref{agile-science10} shows a 1-day
integration of the gamma-ray sky of the Galactic anticenter
region(containing the Crab, Geminga as well as the Vela gamma-ray
pulsars). This field was repeatedly observed with 1-day pointings
with the Crab pulsar position at different angles with respect to
the Instrument axis. The in-orbit calibration phase was
successfully completed at the end of October, 2007 achieving a
good performance for both the gamma-ray and hard X-ray imagers.

The nominal Science Verification Phase observations were
interrupted for specific pointings that were carried out either
for planned multifrequency campaigns (for 3C~279 and 3C~273 in
early July, 2007) or reacting to external alerts due to relevant
optical activity of blazars (3C 454.3 in mid-July, 2007; TXS
0716+714 at the end of October, 2007, W Comae in June, 2008). The
AGILE Cycle-1 program of nominal scientific observations started
on December 1, 2007.


\section{The AGILE Data Center}
\label{ADC}

AGILE science data (about 300 Mbit/orbit) are telemetered from the
satellite to the ASI ground station in Malindi (Kenya) at every
satellite passage (approximately every 95 minutes). A fast ASINET
connection between Malindi and the  Satellite Control Center at
Fucino and then between Fucino and the ASI Science Data Center
(ASDC) ensures the data transmission every orbit.

Scientific data storage, quicklook analysis and management of the
AGILE Guest Observer Program  are carried out at ASDC. After
pre-processing, scientific data (level-1) are corrected for
satellite attitude data and processed by a dedicated software.
Background rejection and photon list determination are the main
outputs of this first stage of processing. Level-2 data are
produced for a full  scientific analysis.

Gamma-ray data generated by the GRID are analyzed by a dedicated
special software producing: (1) sky-maps, (2) energy spectra, (3)
exposure, (4) point-source analysis products, and (5) diffuse
gamma-ray emission. This software is aimed to allow the user to
perform a complete science analysis of specific pointlike
gamma-ray sources or candidates, as well as a timing analysis.
This software is available for the guest observers.

Super-AGILE data are deconvolved and processed to produce 2-D sky
images through a correlation of current and archival data of hard
X-ray sources. Super-AGILE dedicated software produce lightcurves,
spectra and positioning of sources detected in the hard X-ray (18
-- 60 keV).

The AGILE data processing main tasks carried out at the ADC can be
summarized as follows:

\begin{itemize}

\item \textbf{Quicklook Analysis (QLA)} of all gamma-ray and hard
X-ray data, aimed at a fast scientific processing (within a few
hours/1 day depending on source intensity) of all AGILE science
data.

\item \textbf{web-availability of QLA results} to the international community
for alerts and rapid follow-up observations (http://asdc.asi.it,
http://agile.iasf-roma.inaf.it);

\item \textbf{GRB positioning and alerts  through the AGILE Fast
Link}, capable of producing alerts within 10-30 minutes after
trigger;

\item \textbf{standard science analysis } of specific gamma-ray sources open
to the AGILE Guest Observer Program;

\item \textbf{web-availability} of the standard analysis results
of the hard X-ray monitoring program by Super-AGILE.

\end{itemize}

\section{The AGILE Scientific Management}

The Mission management is based on:

\begin{itemize}

\item the AGILE Mission Board (AMB)

\item the AGILE Mission Directorate

\item the AGILE Users Committee

\end{itemize}

Information on the committee membership can be found at
http://asdc.asi.it.

\subsection{The AGILE Mission Board}

The AGILE Mission Board is the executive Board overseeing all
Mission operations and is entitled to make final decisions
regarding all scientific and technical issues.

The AMB is composed by the Principal Investigator, the
co-Principal-Investigator, two ASI-appointed scientists and by the
AGILE Mission Director. The AMB approves the satellite scheduling,
decides on repointings and Mission-of-Opportunity observations,
and decides on issues regarding the AGILE Guest Observer Program.

\subsection{The AGILE Mission Directorate}

The ASI appointed AGILE Mission Director (MD) oversees all
satellite operations, and decides about all technnical issues
regarding the optimal satellite operations, and satellite
pointings.

\subsection{The AGILE Users Committee}

The AGILE Users Committee (AUC) is a 5-member group of scientists
not associated with the AGILE Team. The AUC advices the AGILE
Mission Board on issues regarding the satellite pointing strategy
and data access to the astrophysical community.

\subsection{The AGILE Pointing Program}

The AGILE Pointing Plan is determined according to the Mission
scientific program as proposed by the AGILE Team to the AMB after
the recommendations of the AUC. Solar panel constraints determine
the observability of a given sky region depending on the season.

\subsubsection{Baseline Pointings}

The Cycle-1 of AGILE scientific observations started on December
1, 2007 and will last one year. The Cycle-1 is based on a set of
pre-established baseline pointings approved by the AMB. A detailed
pointing schedule of the AGILE Cycle-1 phase can be found in
http://asdc.asi.it.

\subsubsection{Satellite Repointings}

AGILE can react to transient events of special relevance occurring
outside the current pointing FOV by properly re-pointing the
satellite (always satisfying the solar panel constraints). The
re-pointing has to be approved by the AGILE Mission Board and MD,
and a special procedure is implemented by the MOC to re-orient the
satellite within the shortest possible time. Typical satellite
maneuver timescales are of order of 1-2 orbits.

 For important transients detected within the AGILE-GRID
 and not in the  Super-AGILE FOV,  minor
re-pointings (20-30 degrees) are envisioned to allow
 the coverage of the
gamma-ray transient also by the X-ray imager.

Drastic re-pointings requiring a major re-orientation of the AGILE
satellite are treated as Target-of-Opportunity observations and
are foreseen for events of major scientific relevance detected by
other observatories.

\section{The AGILE Guest Observer Program}

AGILE is a Small  Scientific Mission with a science program open
to the international scientific community. A fraction of the AGILE
gamma-ray data are available to the AGILE Guest Observer Program
(GOP) that is open to the international community on a competitive
basis. The AGILE Cycle-1 GOP started in December, 2007. Details
about the AGILE scientific programme can be found in
http://www.asdc.asi.it.

\section{The AGILE Multiwavelength Program}

The scientific impact of a high-energy Mission such as AGILE
(broad-band energy coverage, very large fields of view) is
greatly increased if an efficient program for fast follow-up
and/or monitoring observations by ground-based and space
instruments is carried out.
 The AGILE Science Program  overlaps and  is
complementary to those of many other high-energy space Missions
(INTEGRAL, RXTE, XMM-Newton, Chandra, SWIFT, {\ch Suzaku }, GLAST)
and ground-based instrumentation (radio telescopes, optical
observatories, TeV observatories).

The AGILE Science Program potentially involves
 a large astronomy and astrophysics community
and emphasizes a quick reaction to transients and a rapid
communication of crucial data.  Past experience has shown that
because of a lack of fast reaction to gamma-ray transients (within
a few hours/days for unidentified sources) many gamma-ray sources
could not be identified. AGILE  takes advantage of the combination
of its gamma-ray and hard X-ray imagers.

 Several
working groups are operational on a variety of scientific topics
including blazars, GRBs, pulsars, and Galactic compact objects.
The AGILE Team together with the ASDC is open to collaborations
with multifrequency observing groups.

\bigskip
\bigskip

 \newpage


 \section*{ACKNOWLEDGMENTS}
\vskip .4cm
 The AGILE program has been developed under the auspices of
the Italian Space Agency with co-participation of the Italian
Institute of Astrophysics (INAF) and of the Italian Institute of
Nuclear Physics (INFN). A very important support to the project
has been provided by CNR and ENEA. The scientific research carried
out for the project has been partially supported under the grants
ASI-I/R/045/04 and ASI-I/089/06/1.  We acknowledge the crucial
programmatic support of the current (G.F.~Bignami) and former ASI
Presidents (S.~De~Julio, S.~Vetrella), as well as of the former
CNR President (L.~Bianco), the former INAF President
(P.~Benvenuti), the former INAF Special Commissioner (S.~De Julio)
and the current INAF (T.~Maccacaro) and INFN (R.~Petronzio)
Presidents.

The ASI Director General (L. De Magistris)
provided a crucial support to the Mission. Several ASI personnel
supported the Mission in different phases and conditions. We
mention here S. Di Pippo, D. Frangipane, P. Caporossi, M. Cosmo,
P. Cecchini,
Q.~Rioli, C.~Musso.

INAF and INFN directorates and President offices offered a very
important support to the Mission. We are particularly grateful to
the Director of the INAF Projects Office (P. Vettolani) and INFN
Directorate member (B. D'Ettorre Piazzoli), and INFN Commission-2
Director (F.~Ronga). We recognize the critical help and unfailing
support offered by the Directors G.~Villa, D.~Maccagni,
R.~Mandolesi,  P.~Ubertini, A.~Vacchi.

A large number of scientists and engineers contributed in a
substantial way at different stages of the project for the success
of the Mission. We mention here the executive Directors
L.~Zucconi, M.~Muscinelli, A.~Beretta, R.~Aceti,  F.~Longoni,
G.~Fuggetta, R.~Cordoni, R.~Starec, and the managers G.~Cafagna
and R.~Terpin. We also thank N. Kociancic, P. Bresciani.
A.~Ercoli-Finzi supported and offered advice at critical phases of
the project. A particular recognition is given to the very
delicate task carried out by G. Grossi.  We also thank the IABG
management (Munich) and in particular A.~Grillenbeck.

 We warmly thank the INFN-LNF staff for invaluable
support and help during the November, 2005, AGILE calibration
campaign. In particular we thank the LNF Director M. Calvetti, and
G. Mazzitelli, P. Valente, M. Preger, P. Raimondi.

Special recognition is to be given to the outstanding technical
and management performance of the ISRO personnel during the AGILE
launch campaign in India. Special thanks are for the ISRO
Sriharikota base Director A. Nair, and to the PSLV C-8 launch
Director N. Narayanamoorthy and his very skilled team. We
acknowledge the collaborative support of the ANTRIX Director K.R.
Sridhara Murthi, and D. Radhakrishnan.

Great support to the Mission has been provided by the CIFS
Scientific Secretariat (G. Ardizzoia), and by the AGILE Team
Scientific Secretariat members, C. Mangili, B. Schena. We
acknowledge the skills and the effective collaboration of the
current AGILE Team Secretariat,  E. Scalise and L. Siciliano.
Personnel of the CNR, INAF and INFN administrations were (and are)
very important for the success of the mission. We thank in
particular the administrations of INAF-IASF Rome, Milan and
Bologna for their support.

Updated documentation on the AGILE Mission can be found at the web
sites http://www.asdc.asi.it, and http://agile.iasf-roma.inaf.it.



\end{document}